\documentclass{emulateapj}
\usepackage{natbib}
\slugcomment{ApJ, in prep.}
\shorttitle{The properties of three ultra-faint dwarfs}
\shortauthors{Sand et al.}

\begin{document}
 \title{Tidal Signatures in the Faintest Milky Way Satellites: The Detailed Properties of Leo~V, Pisces~II and Canes
 Venatici~II }

\author{David J. Sand,$\!$\altaffilmark{1,2,3} Jay Strader,$\!$\altaffilmark{4} Beth Willman,$\!$\altaffilmark{5} Dennis
Zaritsky,$\!$\altaffilmark{6} Brian McLeod,$\!$\altaffilmark{4} Nelson Caldwell,$\!$\altaffilmark{4} Anil Seth,$\!$\altaffilmark{7} Edward
Olszewski $\!$\altaffilmark{6}} \email{dsand@lcogt.net}

\begin{abstract}

We present deep wide-field photometry of three recently discovered faint Milky Way satellites: Leo~V, Pisces~II, and Canes Venatici~II. Our main goals are to study the structure and star formation history of these dwarfs; we also search for signs of tidal disturbance. The three satellites have similar half-light radii ($\sim 60-90$ pc) but a wide range of ellipticities. Both Leo~V and CVn~II show hints of stream-like overdensities at large radii. An analysis of the satellite color-magnitude diagrams shows that all three objects are old ($>$ 10 Gyr) and metal-poor ([Fe/H] $\sim -2$), though neither the models nor the data have sufficient precision to assess when the satellites formed with respect to cosmic reionization. The lack of an observed younger stellar population ($\la 10$ Gyr) possibly sets them apart from the other satellites at Galactocentric distances $\ga 150$ kpc. We present a new compilation of structural data for all Milky Way satellite galaxies and use it to compare the properties of classical dwarfs to the ultra-faints. 
The ellipticity distribution of the two groups is consistent at the $\sim$2-$\sigma$ level.
However, the faintest satellites tend to be more aligned toward the Galactic center, and those satellites with the highest ellipticity ($\ga 0.4$) have orientations ($\Delta \theta_{GC}$) in the range $20^{\circ} \lesssim \Delta \theta_{GC} \lesssim 40^{\circ}$. This latter observation is in rough agreement with predictions from simulations of dwarf galaxies that have lost a significant fraction of their dark matter halos and are being tidally stripped.

\end{abstract}
\keywords{none} 

\altaffiltext{1}{Harvard Center for Astrophysics and Las Cumbres Observatory Global Telescope Network Fellow}
\altaffiltext{2}{Las Cumbres Observatory Global Telescope Network, 6740
Cortona Drive, Suite 102, Santa Barbara, CA 93117, USA}
\altaffiltext{3}{Department of Physics, Broida Hall, University of
California, Santa Barbara, CA 93106, USA}
\altaffiltext{4}{Harvard-Smithsonian Center for Astrophysics, 60 Garden Street, Cambridge MA 02138}
\altaffiltext{5}{Haverford College, Department of Astronomy, 370 Lancaster Avenue, Haverford PA 19041}
\altaffiltext{6}{Steward Observatory, University of Arizona, Tucson, AZ 85721}
\altaffiltext{7}{Department of Physics and Astronomy, University of Utah, Salt Lake City, UT 84112}
\altaffiltext{*}{This paper includes data gathered with the 6.5m Magellan Telescopes located at Las Campanas Observatory, Chile and the Subaru Telescope, which is operated by the National Astronomical Observatory of Japan.}

\section{Introduction}

By detecting slight over-densities of stars with the appropriate color and magnitude in the Sloan Digital Sky Survey (SDSS) catalogs, over 15 new, faint satellites have been discovered around the Milky Way (MW) in the last eight years \citep[][see \citet{willman09} for a recent review]{Will05will1,Willman05umai,Zucker06,Zucker06cvni,Belokurov06,Belokurov07,Irwin07,leov,Belokurov09,Belokurov10,Walsh07booii}.  As a population, these objects are less luminous  (with $-1.5 < M_{V} < -8.6$), and more extended (with $20 \lesssim r_{h} \lesssim 220$ pc) than a typical globular cluster.  Initial studies of these objects indicate that they are predominantly old, metal poor, and dominated by dark matter \citep[e.g.][]{sdsssfh,Kirby08,simongeha}.

Cold Dark Matter (CDM) simulations predict more than an order of magnitude more DM subhalos orbiting the MW than we see as luminous satellites, a discrepancy that increases with increasing simulation resolution \citep[e.g.][]{M99b,Diemand07}. However, connecting simulations with observations is not trivial.  If the CDM picture is correct, then we can test the physics of galaxy formation by comparing the observed numbers and properties of the MW satellites with the predictions of simulations that incorporate such physics \citep[see e.g.][for a recent review]{Kravtsov10}.  On the observational side, an in depth understanding of the new satellites' star formation history (SFH), structure, mass, internal dynamical state and orbital history around the MW are all necessary before we can understand how stars populate the smallest DM halos.

Among the new MW satellites, Canes Venatici II (CVn~II), Leo~V, and Pisces~II are an intriguing triplet with similar properties \citep[see][for their discovery papers, respectively]{Belokurov07,leov,Belokurov10}.  They are all very distant from the MW ($D > 150$ kpc), have similar half light radii ($50 \lesssim r_{h} \lesssim 100$ pc) and are among the faintest objects (with $-4.0 \lesssim M_{V} \lesssim -5.0$) detectable in the SDSS at that distance  \citep{Koposov08,Walsh09}.    Based on spectroscopic results, CVn~II appears to be dark matter dominated with a $M/L_{V}=360^{+380}_{-180}$ \citep{Wolf10} and is metal poor, with $\langle$[Fe/H]$\rangle=-2.19$ \citep{Kirby08}.  On the other hand, neither Leo~V nor Pisces~II have robust mass estimates, although the kinematic study of \citet{Walker09} suggests that Leo~V might be dark matter dominated.  Leo~V bears several hints that it might have been tidally stripped, including a spectroscopic [Fe/H] that is higher than MW satellites of similar luminosity, and more compatible with brighter objects.  Leo~V also shows signs of being disturbed, with an apparently extended blue horizontal branch distribution \citep{leov} and kinematic members more then 10 half light radii away from the satellite center \citep{Walker09}.  There have also been suggestions that Leo~V and the nearby satellite Leo~IV are connected  by a `stellar bridge', and share a common origin \citep[][but see \citet{Sandleoiv}]{leoivleov}.  Pisces~II has had no published spectroscopic follow up to date.

Their stellar populations, based on earlier work, appears to be uniformly old ($\gtrsim$ 10 Gyr) and metal poor ([Fe/H]$\lesssim-2$), with no obvious sign of more recent star formation.  If confirmed, then these three satellites have qualitatively different formation histories than other similarly distant MW satellites.  Among the eight classical dSphs, the four with the largest distance from the MW ($D>90$ kpc) all have extended star formation histories (SFHs), with significant amounts of star formation occurring within the last 10 Gyr \citep{Dolphin05}.  The other two faint MW satellites with $D > 150$ kpc, Leo~IV and CVn~I, both have signs of more recent star formation based on an excess number of blue plume stars which appear clumped and segregated within the main body of the satellite \citep[][but see \citet{Okamoto12}]{Martin08,Sandleoiv}.

Motivated by all of the above, we obtained deep, wide-field ($\sim25'\times25'$) photometry of Leo~V, CVn~II and Pisces~II  on $\sim$6-8 meter class telescopes, with the goal of ascertaining their structure and star formation history.   An outline of the paper follows.  In Section~\ref{sec:datareduce}, we describe the observations, data reduction, photometry and final point source catalogs of our objects.  In Section~\ref{sec:dist} we measure the distance to our satellites (when not already established with RR~Lyrae stars in the literature) via the magnitude of their blue horizontal branch (BHB) stars.  Section~\ref{sec:structure} details the structural properties of our three satellites including both a parameterized fit to their ridgeline stars and a matched-filter search for signs of extended/disturbed structure.  Section~\ref{sec:sfh} presents a qualitative analysis of each satellite's SFH, and a measurement of their luminosity.  We then take a step back in \S~\ref{sec:discuss} to place the three objects in the context of the new MW satellite population as a whole.  We compare the structural properties of the new satellites with the classical dSphs, and search for signs of tidal disturbance based on their alignment with the vector to the MW center.  Finally, we summarize and conclude in Section~\ref{sec:conclude}.  Throughout this work, we use the terms `ultra-faint satellites', `post-SDSS satellites' and `new MW satellites' interchangeably to represent the MW satellites discovered via matched-filter techniques of the SDSS point source catalogs. 

\section{Observations \& Data Reduction} \label{sec:datareduce}

Here we describe the observations, data reduction and photometry of Leo~V, Pisces~II, and CVn~II used in this paper.  At the end of this section, we present our full Leo~V, Pisces~II and CVn~II photometry catalogs.  A summary of the observations is in Table~\ref{table:observations}.  

We observed Leo~V on 2010 April 13 (UT) and Pisces~II on 2010 Oct 3 (UT) with Megacam \citep{megacam}
on the f/5 focus at the Magellan Clay telescope in the $g$ and $r$ bands.  Magellan/Megacam has 36 CCDs,
each with $2048 \times 4608$ pixels at 0\farcs08/pixel (which were
binned $2 \times 2$), for a total field of view (FOV) of $\sim$24'$\times$24'.
We reduced the data identically to \citet{Sandleoiv} using the Megacam
pipeline developed at the Harvard-Smithsonian Center for Astrophysics by
M. Conroy, J. Roll and B. McLeod.  

Subaru/Suprimecam data of CVn~II comes from the SMOKA science archive\footnote{http://smoka.nao.ac.jp/}, which archives the public data of the Subaru Telescope.  While other data on CVn~II are available, we retrieved the deep $V$ and $I$ imaging observed by Okamoto and collaborators \citep[see][for their presentation of their CVn~II data]{Okamoto12}, as summarized in Table~\ref{table:observations}.  We have reduced the archival Suprimecam \citep{suprimecam} data of CVn~II in a standard way, performing bias subtraction and flat fielding with the provided calibration frames.  Cosmic-ray rejection of each individual exposure was done with the {\sc lacosmic} task \citep{vandokkum01}.  An astrometric solution was found via {\sc SCAMP} using the SDSS-DR7 catalog.  Once a good astrometric solution was found, the image resampling and co-addition software {\sc SWarp}\footnote{http://astromatic.iap.fr/software/swarp/} was employed (with the {\sc lanczos3} interpolation function) using a weighted average of the individual frames to make our final image stack.  Note that we have not included one corner CCD (DET-ID 0) into our final image stacks due to poor charge transfer efficiency at the time of the observations.

We performed stellar photometry on the final image stacks using a methodology identical to that in our previous works \citep{sandherc,Sandleoiv} with the command line version of the {\sc DAOPHOTII/Allstar} package \citep{Stetson94}.  Here we briefly reiterate.  We
allowed for a quadratically varying point spread function (PSF) across the field when
determining our model PSF and ran {\sc Allstar} in two passes -- once
on the final stacked image and then again on the image with the first
round's stars subtracted, in order to recover fainter sources.  We
culled our {\sc Allstar} catalogs of outliers in $\chi^{2}$ versus
magnitude, magnitude error versus magnitude and sharpness versus
magnitude space to remove objects that were not point sources.  We
positionally matched our source catalogs derived from different filters with a
maximum match radius of 0\farcs5, only keeping those point sources
detected in both bands in our final catalog.

We converted instrumental magnitudes into the SDSS photometric system using stars in common with SDSS-DR7, as in \citet{Sandleoiv}, which included a color term.  For our $V$ and $I$ band observations of CVn~II, we used the filter transformations of \citet{Jordi06} to convert from SDSS catalog magnitudes to $V$ and $I$ bands.  Slight residual zeropoint gradients across the FOV were fit to a quadratic function and corrected for \citep[see also][]{Saha10}, resulting in a final overall scatter about the best fit zeropoint of $\delta \lesssim 0.05$ mag in all of our photometric bands.

We performed a series of artificial star tests to calculate our photometric errors and completeness as a function of magnitude and color for each of our fields.  Artificial stars were placed into our images on a regular grid (10 to 20 times the image FWHM), with the DAOPHOT routine ADDSTAR.  Ten iterations were performed on each field, yielding between $\sim6\times10^{5}$ and $\sim8\times10^{5}$ implanted artificial stars.  The $r$ ($I$) magnitude for a given artificial star was drawn randomly from 18 to 29 mag, with an exponentially increasing probability toward fainter magnitudes.  The $g-r$ ($V-I$) color is then randomly assigned over the range ($-0.5$,1.5) with equal probability.  These artificial star frames were then run through the same photometry pipeline as the unaltered science frames, applying the same $\chi^{2}$, sharpness and magnitude-error cuts.  The 50\% and 90\% completeness for each of our fields and imaging bands is summarized in Table~\ref{table:observations}.

\subsection{Final Catalogs and Color Magnitude Diagrams}

We present our full Leo~V, Pisces~II and CVn~II catalogs in Tables 2-4.  Each table includes the calibrated magnitudes (uncorrected for extinction) with their uncertainty, along with the Galactic extinction values derived for each star \citep{Schlegel98}.  We also note whether the star was taken from the SDSS catalog rather than our Magellan or Subaru photometry, as was done for objects near or brighter than the saturation limit of the observations.  All magnitudes reported in the remainder of this paper will be corrected for Galactic extinction.

\begin{figure*}
\begin{center}
\mbox{ \epsfysize=4.0cm \epsfbox{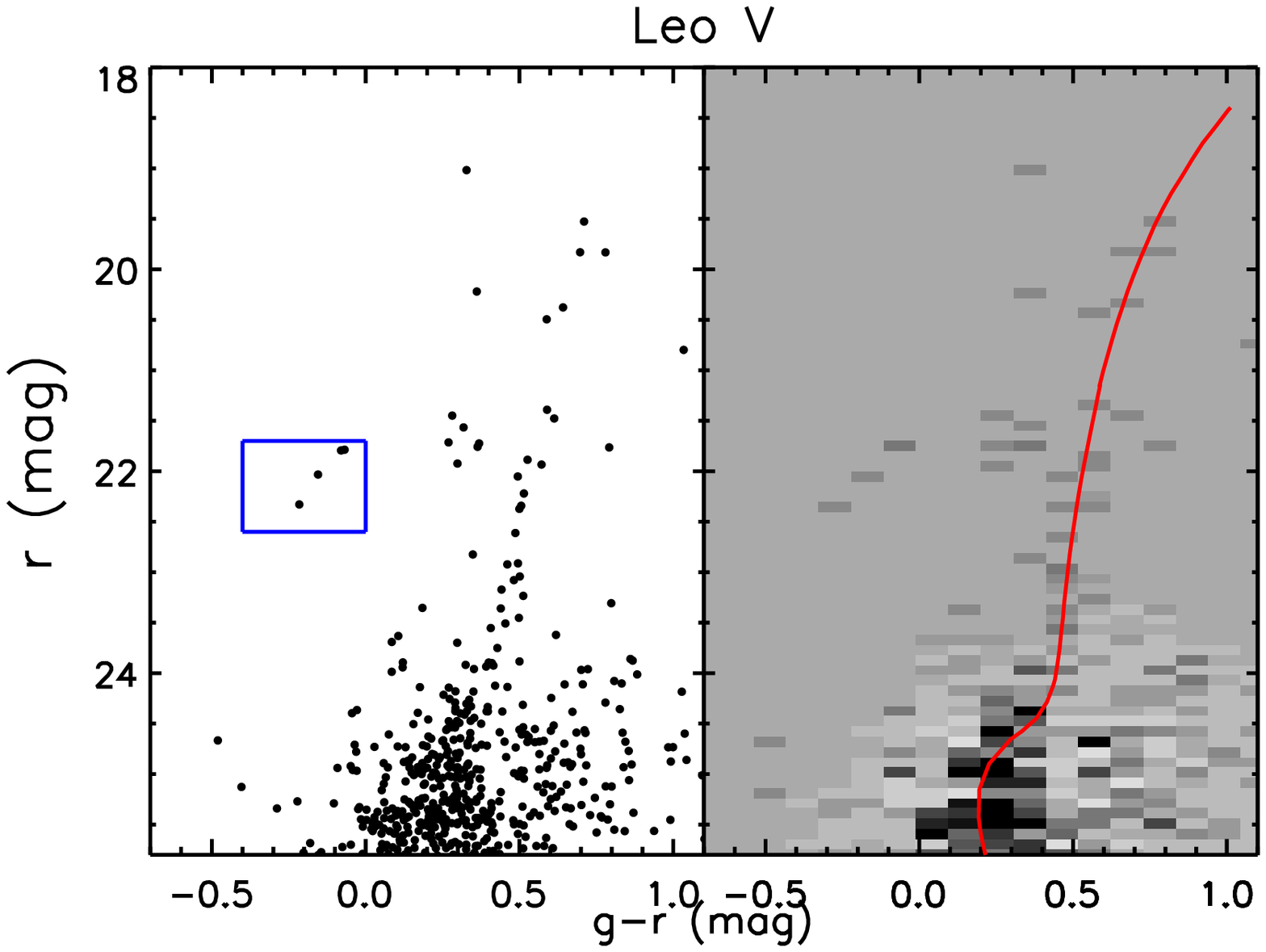} \epsfysize=4.0cm
\epsfbox{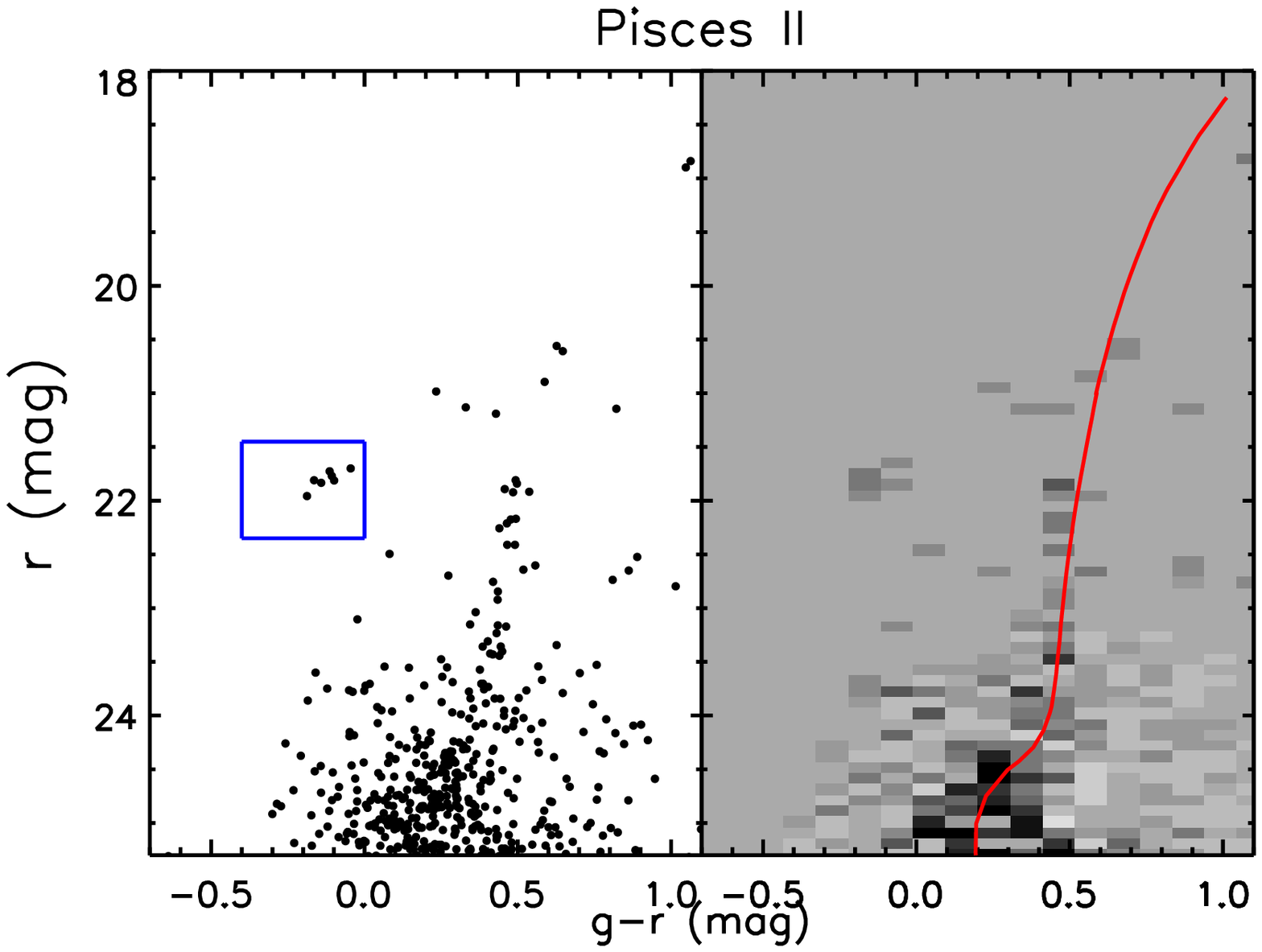} \epsfysize=4.0cm \epsfbox{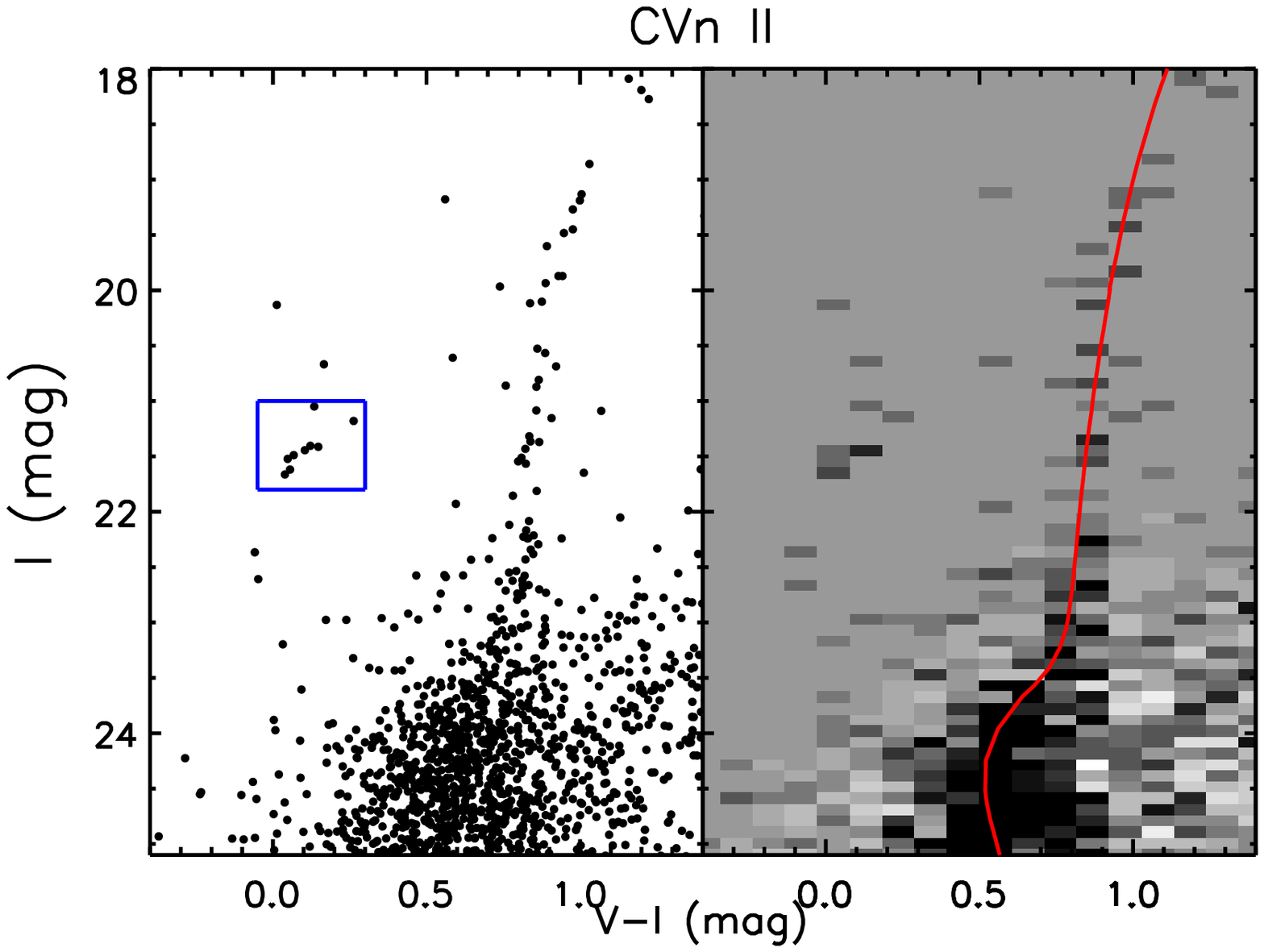}} 
\caption{Color magnitude diagrams of our three satellites -- Leo~V (Left), Pisces~II (Middle), and CVn~II (Right) --  utilizing data within two half light radii (see \S~\ref{sec:paramfit}).  Each satellite's figure is split into two panels.  On the left are the raw CMDs.  The inset box shows our initial selection region for BHB stars, which we will use as our starting point for studying their spatial distribution in \S~\ref{sec:extend}.  On the right is overplotted
 a theoretical isochrone from Girardi et al. (2004) with [Fe/H]=$-2.0$ and 13.5 Gyr age.       \label{fig:CMDs}}
\end{center}
\end{figure*}

The color magnitude diagram (CMD) of each of our satellites is presented in Figure~\ref{fig:CMDs}.  Plotted in the left panel are all stars within two half-light radii (as determined in \S~\ref{sec:paramfit}), while the right panel is a Hess diagram of the same region with a scaled background subtracted, using stars located outside a radius of 8 arcminutes.  We highlight possible stellar populations for each of our satellites in both panels.  In the right panels of each satellites' CMD we plot a theoretical isochrone from \citet{Girardi04}, with [Fe/H]=$-2.0$ and age of 13.5 Gyr, using the satellite distances as determined in \S~\ref{sec:dist}.  The solid box in the left panels in Figure~\ref{fig:CMDs} denotes our initial blue horizontal branch (BHB) star selection region in the CMDs.  We study the BHB spatial distribution in each dwarf in \S~\ref{sec:extend}.

\section{Satellite Distances}\label{sec:dist}

The distance to each of the satellites is necessary for deriving their physical size and placing them in context with respect to the MW satellite population as a whole.

We constrain the distance to Pisces II and Leo V using the luminosity of their horizontal branch (HB) sequence in the following way.  First, two fiducial HB star sequences were constructed using SDSS photometry;  one from the globular cluster (GC) M92 and the other from the union of M3 and M13. For clarity in what follows, we assume distance moduli of $\mu=14.75\pm0.1$, $15.14\pm0.2$ and $14.31\pm0.1$ mag for M92 \citep{Kraft03}, M3 \citep{Cho05} and M13 \citep{Kraft03}, respectively.  The uncertainty in distance modulus for M3 was taken directly from the analysis of \citet{Cho05}, while that for M92 and M13 are estimated based on their agreement with the independent distance measurement of \citet{Vandenberg00}.  We note that our value for M92's distance is $\sim$0.15 mag different from that in the Harris GC catalog \citep[][who report a distance to M92 of 8.3 kpc, or $\mu$=14.6 mag]{Harris96}.  We take literature values of [Fe/H]=$-$2.4 and E(B-V)=0.022 mag for M92,  and [Fe/H]=$-$1.5 and E(B-V)=0.013 mag for M3 and [Fe/H]=$-$1.6 and E(B-V)=0.017 for M13 \citep{Kraft03}.

We then gathered HB stars for Leo~V and Pisces~II, taking stars out to  2.5$r_h$ (see \S~\ref{sec:paramfit}), as this is roughly the distance at which there were clear members with no offset from the HB sequence.  We fit to both of the GC fiducial HB sequences by minimizing the sum of the squares of the difference between the data and the fiducial, and summarize the results in Table~\ref{table:distances}.  Given the relation between metallicity and HB luminosity \citep[see e.g.][for a recent review]{Catelan09}, we also determined distances assuming a nominal [Fe/H]=$-$2.0, given that $M_{V,HB} \propto 0.2\times$[Fe/H], and using the M3/13-derived distance as a baseline.  In this case, we added a further uncertainty of 0.05 mag to represent the uncertainty in the $M_{V,HB} - [Fe/H]$ relation.  We note that Leo~V has a spectroscopic [Fe/H]=$-2.0\pm0.2$ \citep{leovspec}, while Pisces~II has no published spectroscopic metallicity.

Uncertainties are calculated via jackknife resampling, which accounts for both the finite number of stars in each satellite, and the possibility of occasional interloper stars.  The uncertainties associated with our calibration to the SDSS photometric system ($\lesssim0.01$ mag), our globular cluster fiducial distance uncertainties, and reddening ($\sim$0.03 mag) were added in quadrature to produce our final quoted uncertainty.  The systematic uncertainty associated with the metallicity-luminosity relation of HB stars -- which we estimate to be $\sim$0.05 mag -- was added to the uncertainty in the distance derived via our M3/M13 translation to [Fe/H]=$-$2.0.

Our M92-derived distance to Leo~V is offset from that previously reported using the same cluster as a fiducial in the discovery work of \citet{leov}, who found D=178 kpc.  We suspect that this difference is due to the ambiguity in M92's distance rather than the Leo~V photometry itself, although this cannot be definitively tracked down since it is unclear what distance to M92 Belokurov et al. used.  
The distance to Pisces~II determined in the discovery paper of \citet{Belokurov10}, D$\sim$182 kpc, is consistent with our measurements to that satellite.

Given the ambiguity of M92's distance, the lack of a spectroscopic [Fe/H] for Pisces~II, and the reasonable consistency of their stellar populations with the [Fe/H]=$-$2.0 isochrones (see \S~\ref{sec:sfh}), we utilize our [Fe/H]=$-$2.0 distance measurements for Leo~V and Pisces~II for the remainder of this work.

To lend credence to our technique, we have utilized the HB data of Leo~IV from \citet{Sandleoiv}, and measured D=158$\pm$12 kpc for a HB with [Fe/H]=$-$2.0, which is in excellent agreement with the RR~Lyrae distance, D=154$\pm5$ kpc, as determined by \citet{Moretti09}.  Given this, and a lack of globular cluster fiducials in the $V$ and $I$ band (the photometric bands of our CVn~II data), we directly adopt the RR~Lyrae distance of CVn~II measured by \citet{Greco08}
-- $\mu=21.02\pm0.06$ mag ($160\pm7$ kpc) -- which matches the old stellar population isochrones in
Figure~\ref{fig:CMDs}.

\section{Structural Properties}\label{sec:structure}

We split our structural analysis into two parts.  First, we fit parameterized models to the two dimensional density profile of our satellites in order to measure their basic structural properties such as half light radius ($r_{h}$) and ellipticity.   From there, we search for signs of extended structure, such as tidal streams, around each satellite using a matched-filter technique.

\subsection{Parameterized Fits}\label{sec:paramfit}

As in previous work on the classical and ultra-faint MW satellites, we fit standard parameterized models -- the Plummer and exponential distribution -- to the surface density profile of the three objects in the present study.  While the observed MW satellites have a complexity and non-uniformity that can not be characterized with parameterized models, it is nonetheless important to quantify their structure in a consistent way for comparison with previous results.  We note that we do not report King profile parameters for the three objects in this study, as we have in previous work -- due to the small number of stars in each satellite and the additional free parameter in the King model, we did not obtain reliable results (in particular for Leo~V and Pisces~II).

We use a maximum likelihood (ML) technique for constraining structural parameters (based on the recipe of Martin et al. 2008), identical to that done in \citet{Sandleoiv}.  Our slightly modified technique is robust to non-rectangular field of view geometries because all integrals are calculated via Monte Carlo integration -- for instance, Eqn. 5 from \citet{sdssstruct}.  Both the exponential and Plummer profiles have the same set of free parameters -- ($\alpha_{0}$, $\delta_{0}$, $\theta$, $\epsilon$, $r_{half}$, $\Sigma_b$). In order, these include the central position, $\alpha_{0}$ and $\delta_{0}$, position angle (PA; $\theta$), ellipticity ($\epsilon$), half light radius ($r_{half}$) and background surface density ($\Sigma_{b}$).  Uncertainties on structural parameters are determined through 1000 bootstrap resamples, from which a standard deviation is calculated.  
The stars selected for this analysis are those consistent with the ridgeline (red giant branch, subgiant branch and the main sequence) of each satellite.  Briefly, we use a [Fe/H]=$-2.0$, 13.5 Gyr old theoretical isochrone ridgeline \citep{Girardi04}, for the satellite distance modulus determined in \S~\ref{sec:dist}, and placed two selection boundaries at a minimum of 0.1 mag on either side along the $g-r$ axis.  These selection regions are increased to match the typical $g-r$ color uncertainty at a given $r$ magnitude when it exceeds 0.1 mag, as determined via our artificial star tests.  The choice of ridgeline is not crucial, because old stellar populations in the expected metallicity range all fall within $\sim$0.1 mag of the ridgeline.  We impose a faint magnitude limit corresponding to our 50\% completeness limit reported in Table~\ref{table:observations}.  Each selection region was visually checked to verify that stars consistent with the visible ridgeline of each satellite were included in this analysis.  Note that HB stars are not included in this technique, although we do discuss their spatial distribution further in \S~\ref{sec:extend}.

\begin{figure*}
\begin{center}
\mbox{ \epsfysize=4.2cm \epsfbox{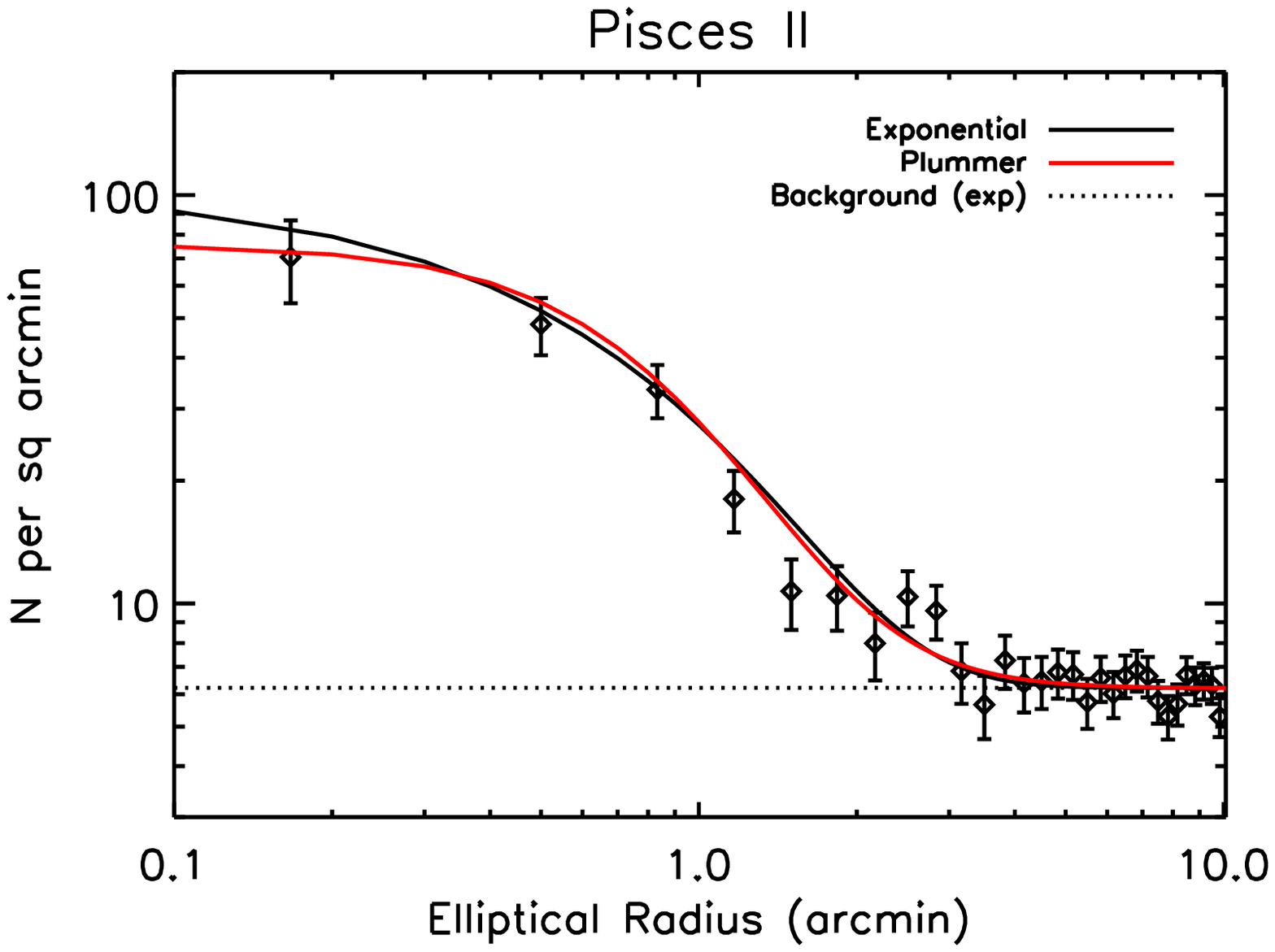} \epsfysize=4.2cm
\epsfbox{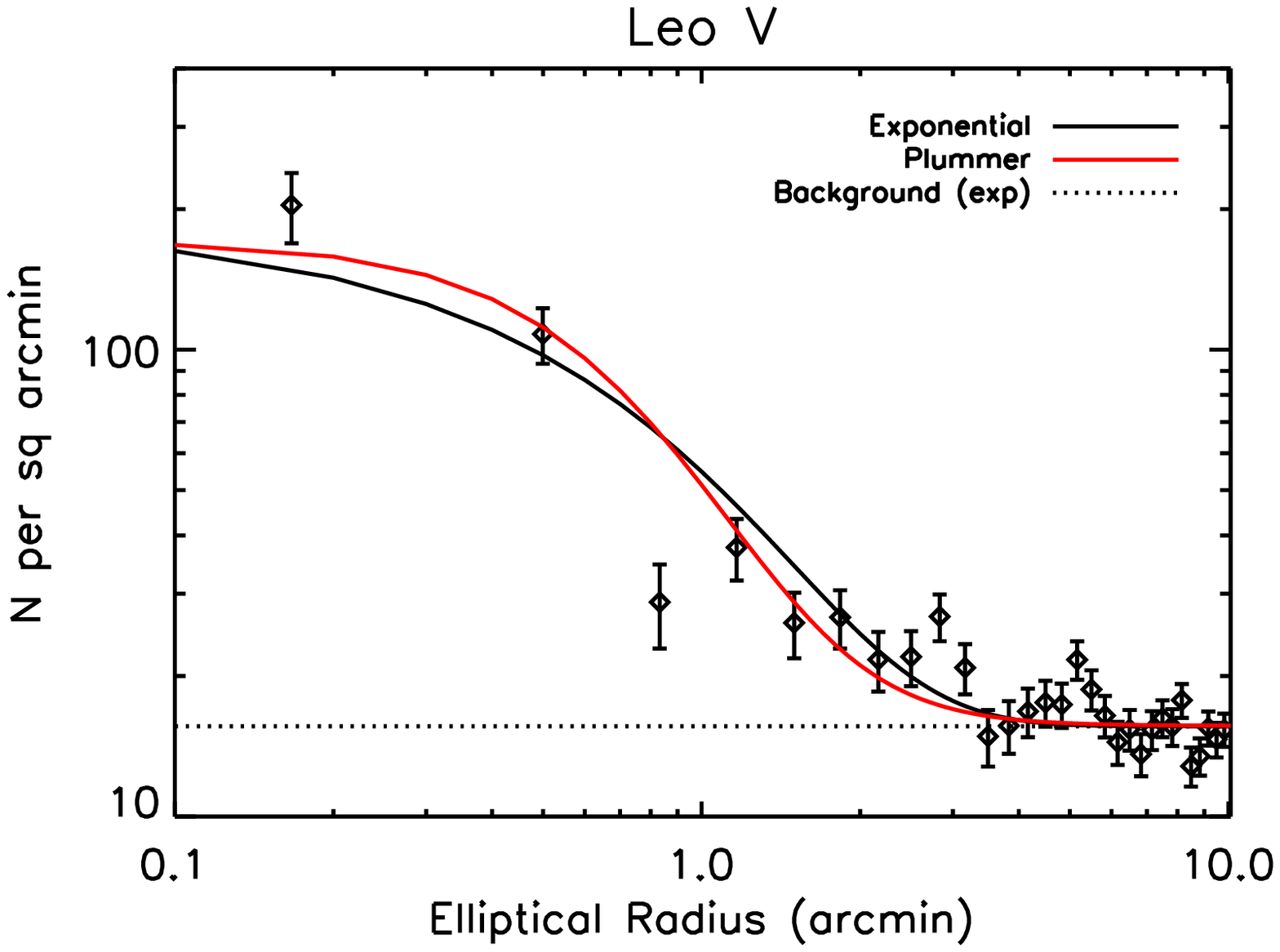} \epsfysize=4.2cm \epsfbox{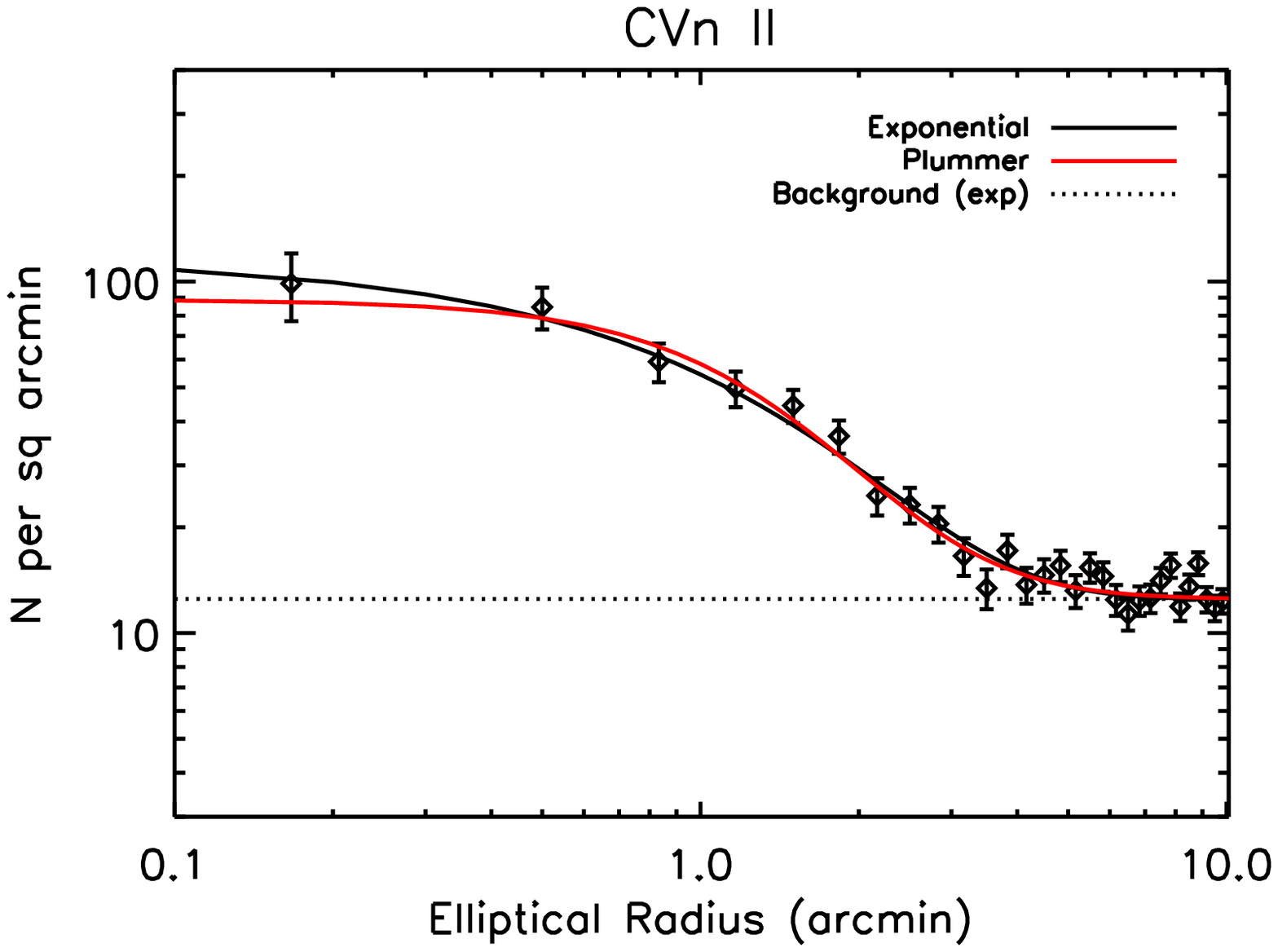}} 

\caption{Stellar profiles of Leo~V, Pisces~II and CVn~II, where the data points are the binned star counts for stars consistent with the CMD ridgeline of each object.  The plotted lines show the best-fit one-dimensional exponential and Plummer profiles, found via our ML analysis in \S~\ref{sec:paramfit}.  The dotted line shows the background surface density determined for our exponential profile fit.  Note that in deriving our profile fits, we are not fitting to the binned data, but directly to the two-dimensional stellar distribution. \label{fig:surfdens} }
\end{center}
\end{figure*}

\begin{figure*}
\begin{center}
\mbox{ \epsfysize=4.2cm \epsfbox{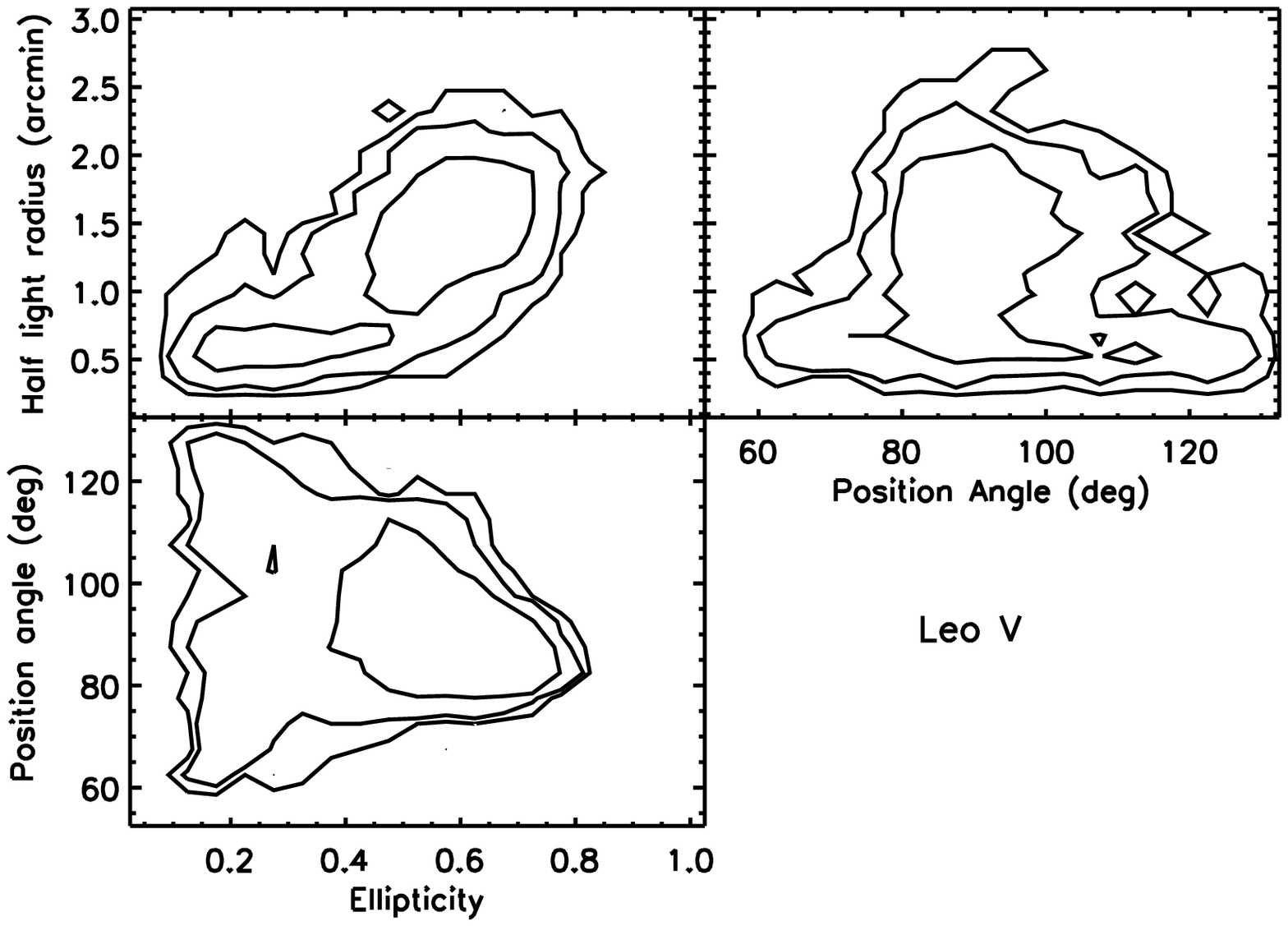} \epsfysize=4.2cm
\epsfbox{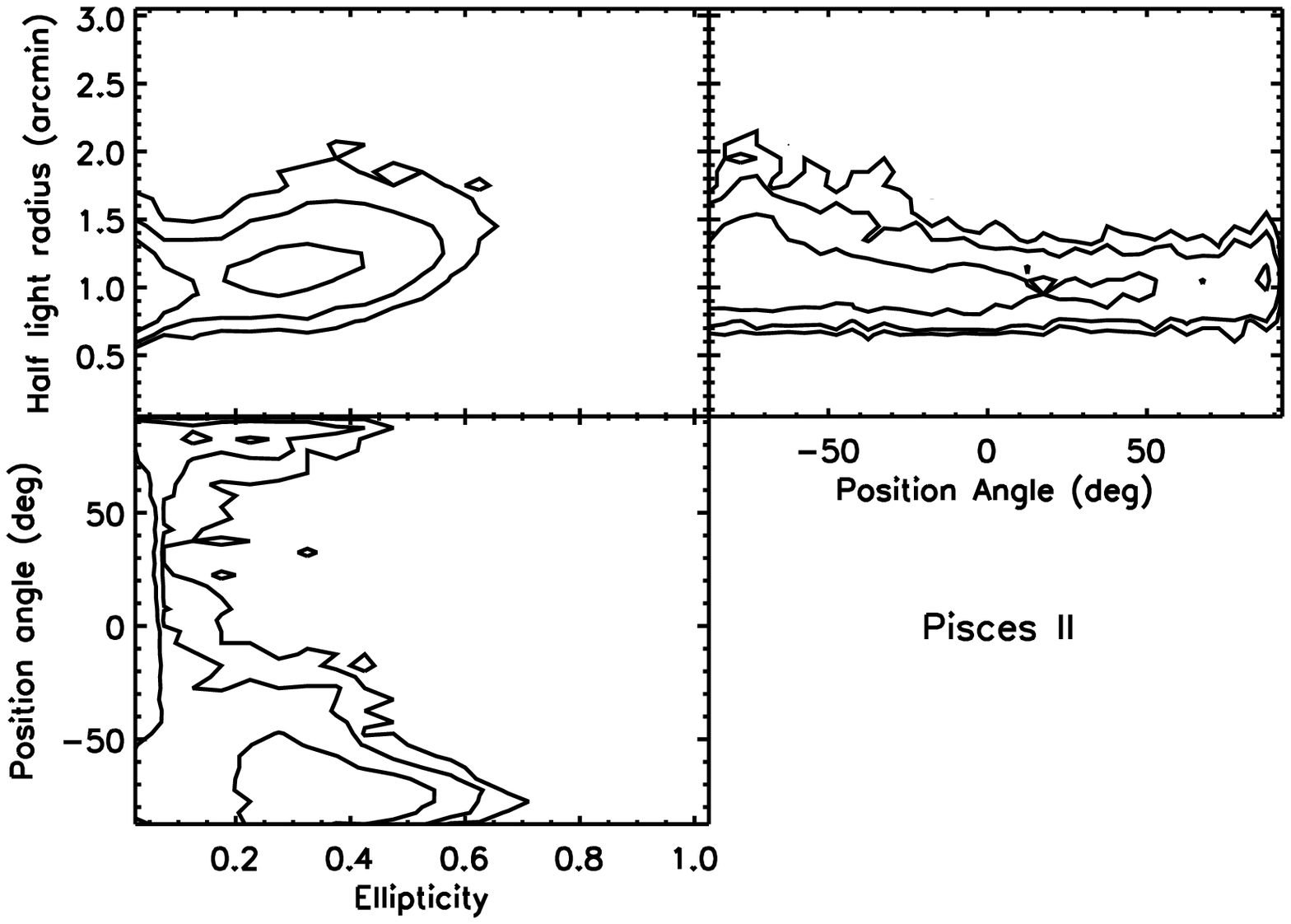} \epsfysize=4.2cm \epsfbox{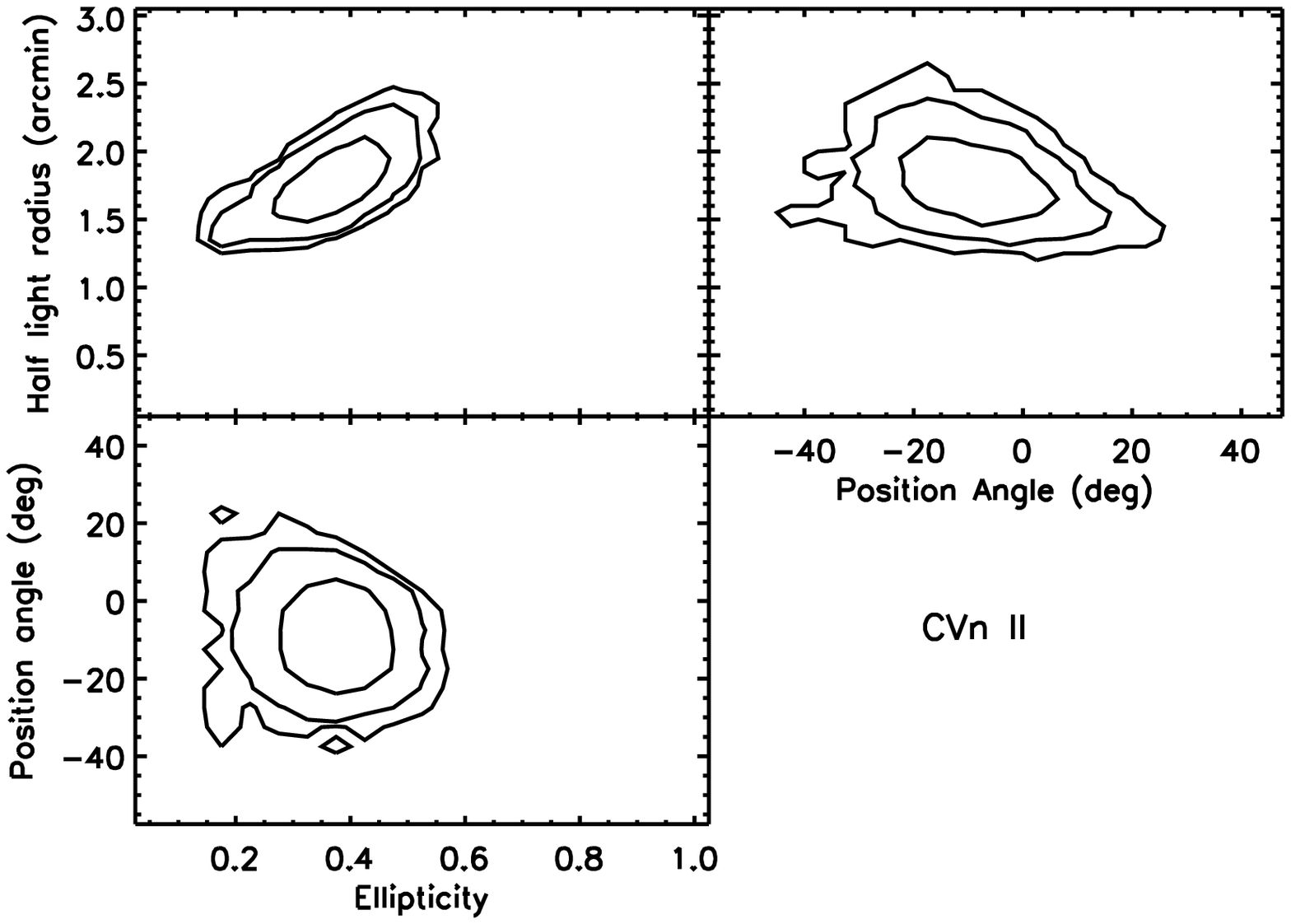}} 

\caption{Two-dimensional, marginalized confidence contours (corresponding to the 68\%, 95\% and 99\% confidence limits) for the half light radius, ellipticity and position angle for the three dwarfs in our study.  These contours are for the exponential profile fit.  \label{fig:contours} }
\end{center}
\end{figure*}

Our results are presented in Table~\ref{table:paramfits}.  We also show one dimensional stellar radial profiles corresponding to our best-fit parameters, along with our binned data, in Figure~\ref{fig:surfdens}.  Despite the fact that we fit structural parameters to the two dimensional distribution of satellite stars, the one-dimensional representation of the fits show satisfactory, but imperfect, agreement.  As mentioned earlier, our parameterized models should not be expected to be excellent descriptions of the satellites' potentially complex structure.  Note that the exponential profile fit to Pisces~II did not result in a well-defined ellipticity, and thus major axis position angle, although the Plummer profile fit was able to constrain these quantities.  Similarly, the half light radius of Leo~V, with either parameterization, had a large ($\sim$50\%) uncertainty.  To illustrate some of the key parameter degeneracies for the exponential profile fit, we show the two-dimensional, marginalized confidence contours for the half light radius, ellipticity and position angle for the three dwarfs in our study in Figure~\ref{fig:contours}.

Recently, \citet{Munoz11} presented a suite of simulations of low luminosity MW satellites under different observing conditions to determine the dataset quality necessary to measure accurate structural parameters.  In particular, in order to get a $\sim$10\% measurement of the half light radius, they suggested a field of view at least three times that of the half light radius being measured, greater than 1000 stars in the total sample, and a central density contrast of 20 over the background.  The data presented in this paper has a central density contrast of $\sim$10 (see Figure~\ref{fig:surfdens}), and this may be responsible for our inability to measure half light radii to $\sim$10\%.


Overall, our results are in agreement with those in the literature where there is overlap.  We measure CVn~II to be marginally rounder ($\epsilon=0.39\pm0.07$) than the structural analysis of \citet{sdssstruct} performed on the shallower SDSS discovery data ($\epsilon=0.52^{+0.10}_{-0.11}$), but otherwise measure similar parameter values considering the uncertainties.  CVn~II was studied with nearly identical Subaru data as our own by \citet{Okamoto12}, who found the same half light radius as we do, but a smaller ellipticity, $\epsilon$=0.23 (with no associated uncertainty).  This can be attributed to the fact that Okamoto et al. measured their ellipticity after binning and smoothing the data, which is known to systematically lower the resulting ellipticity measurement \citep{sdssstruct}.  Our derived half light radius is inconsistent with the deep data obtained for the RR~Lyrae study done by \citet{Greco08}, who found $\sim150$ pc, although their measurement is inconsistent with other literature values as well.  It is unclear where this discrepancy originates.  Leo~V and Pisces~II have been discovered since the homogenous analysis of \citet{sdssstruct}, but our results are broadly consistent with the discovery data of each object \citep{leov,Belokurov10}.

\subsection{Extended Structure Search}\label{sec:extend}

Given the intriguing structural properties of the new MW satellites as a population \citep[e.g.][]{sdssstruct}, and hints that some of the satellites may be tidally disturbed \citep[e.g.][]{sandherc,Munoz10}, it is worth searching for signs of extended structure, such as streams or other extensions, within our data. We use a matched-filter technique identical in spirit to that of \citet{Rockosi02}, with a well understood `signal' CMD and `contamination' CMD.  We refer the reader to \citet{Rockosi02} for the details of the algorithm.  The matched-filter technique, in contrast to simpler star-counting density maps, gives higher weights to stars which have relatively low contamination, such as BHB stars.

To be successful, we must have high-fidelity signal and contamination CMDs.  For the purposes of the matched-filter technique, we bin all of our CMDs into $0.15\times0.15$ color-magnitude bins in what follows.  Rather than use stars from the central regions of each satellite for our signal CMD (which will be sparsely populated and will contain background/foreground stars), we use theoretical isochrones which match our observed satellite CMD and are convolved  with our well-understood completeness and uncertainty functions.  We take our cue from the qualitative stellar population analysis presented in \S~\ref{sec:stellpop}.  To be specific, we use the {\it testpop} program in the StarFISH software suite; a set of programs designed for fitting the star formation history of observed stellar populations \citep{starfish}.  The {\it testpop} program will take a theoretical isochrone set, convolve it with the photometric completeness and uncertainties of your observations, and produce a realistic, `observed' CMD. For each satellite, we create realistic CMDs populated with 50000 stars as our signal CMD, using a single stellar population with [Fe/H]=$-2.0$  and an age of 13.5 Gyr from the results of \citet{Girardi04}.  We show in \S~\ref{sec:stellpop} that this stellar population is consistent with that seen in our three satellites.

For our contamination, background CMD we use all stars at radii larger than 8 arcmin from our satellite galaxies.  Given the small size of our satellites, with half light radii $\lesssim$2 arcmin, this should yield a relatively pure background CMD, unless of course our satellites have low density streams or extensions out to these radii.  To guard against this, we repeated our analysis with background CMDs taken only from each of the four separate quadrants of our field of view.  These spatially distinct background CMDs did not change our final satellite maps significantly, suggesting that there are no streams which we are washing out by including signal into our background contamination filter.

\begin{figure*}
\begin{center}
\mbox{ \epsfysize=6.0cm \epsfbox{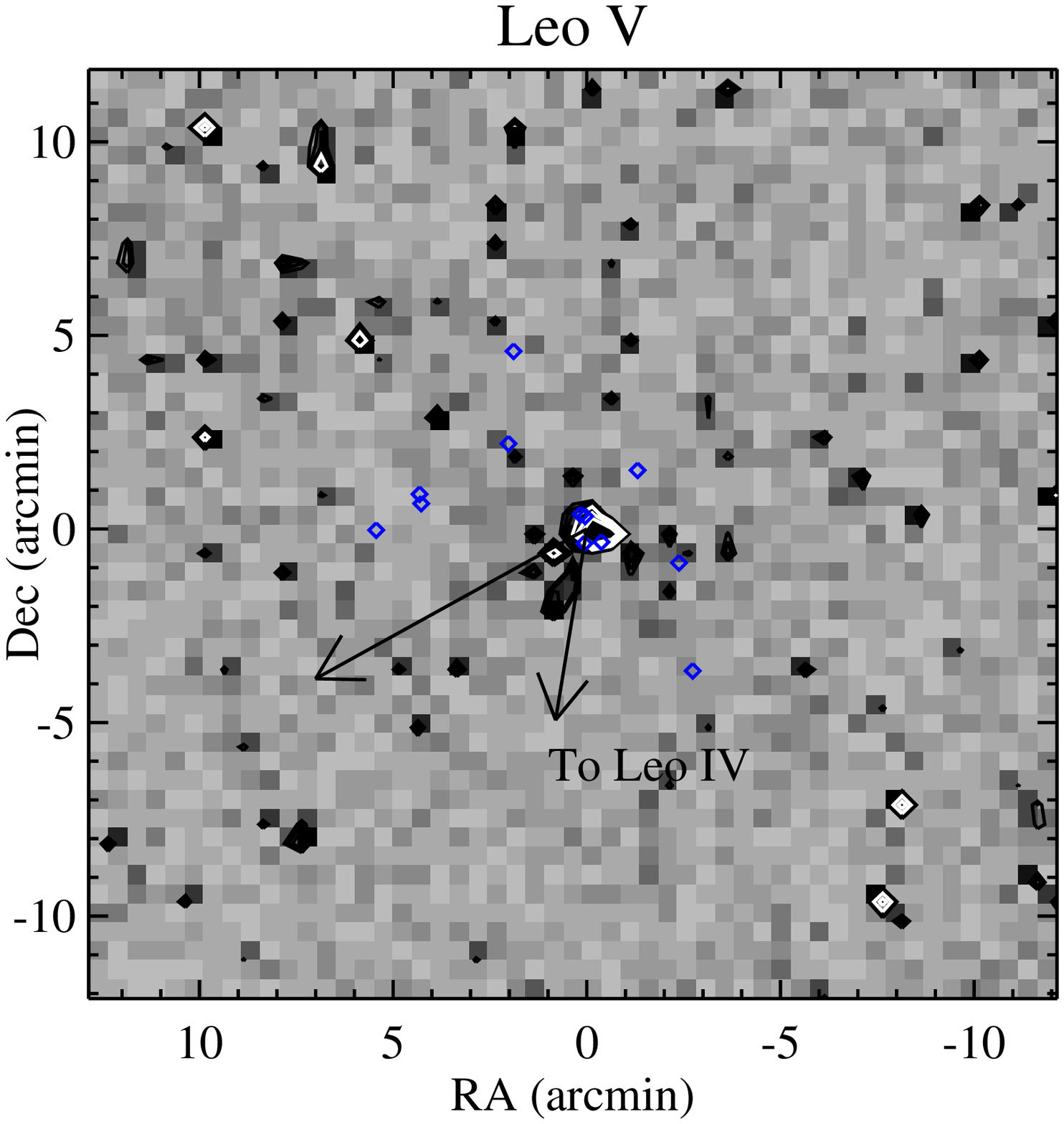} \epsfysize=6.0cm
\epsfbox{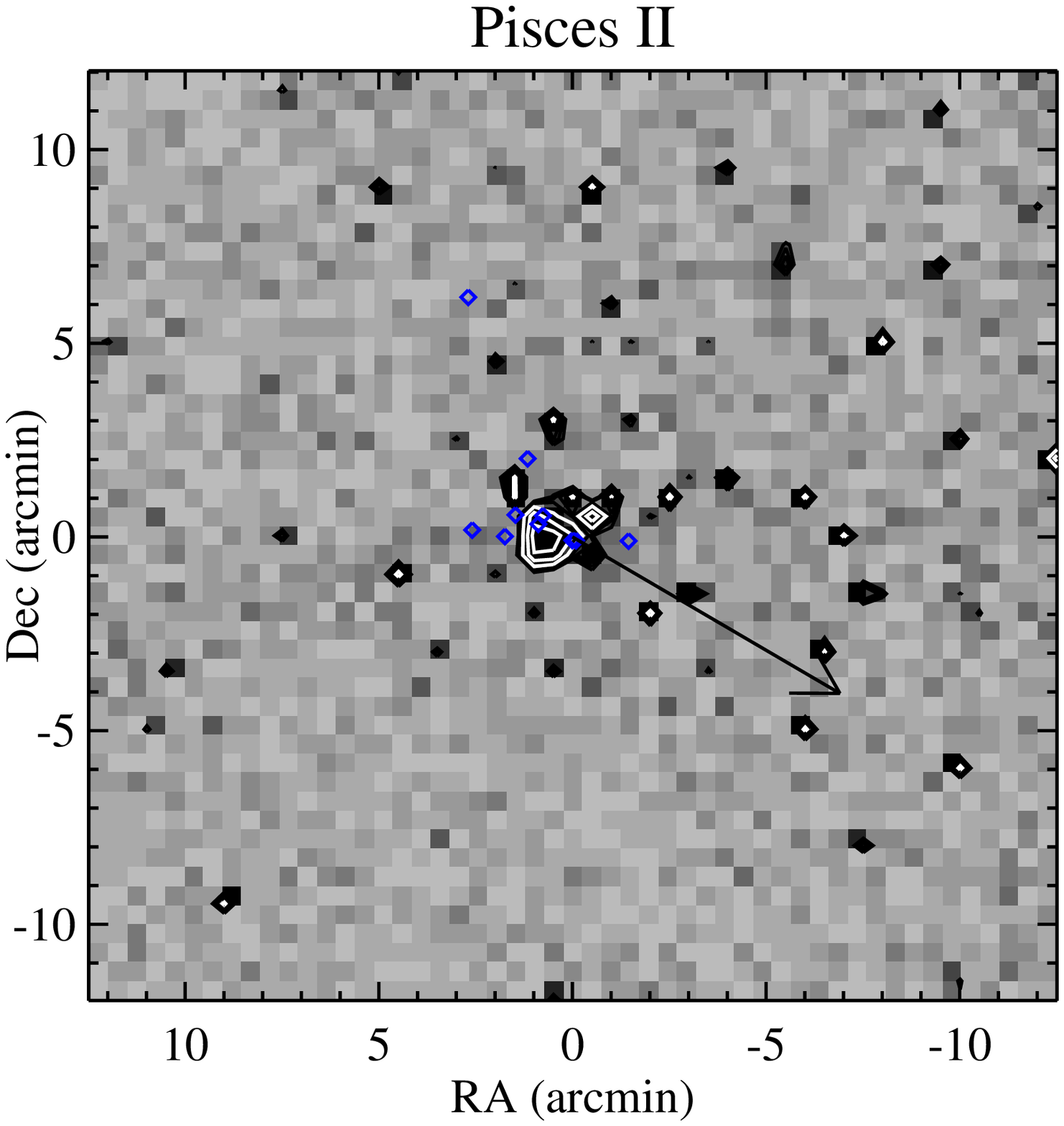} \epsfysize=7.0cm \epsfbox{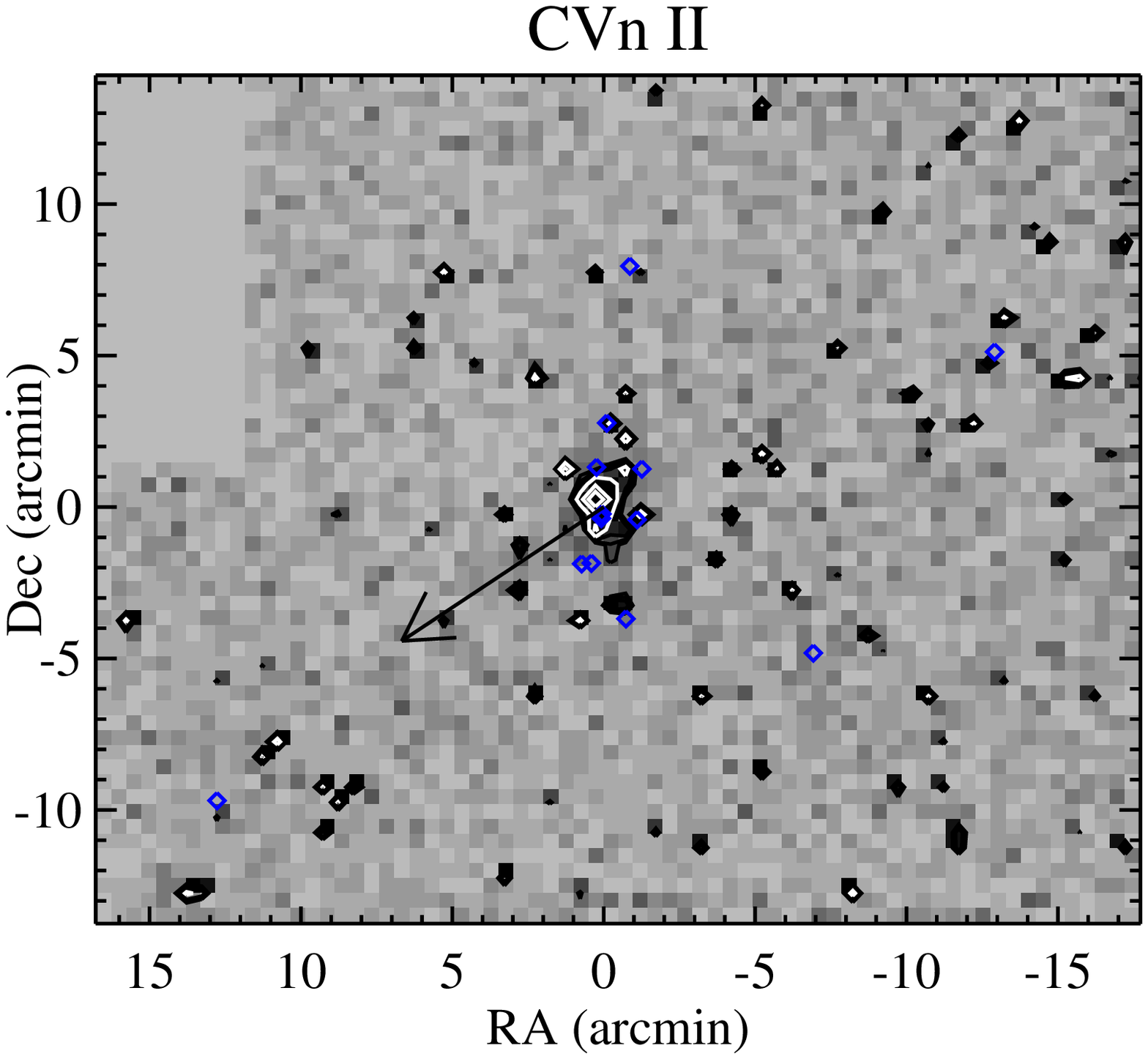}} 
\mbox{ \epsfysize=6.0cm \epsfbox{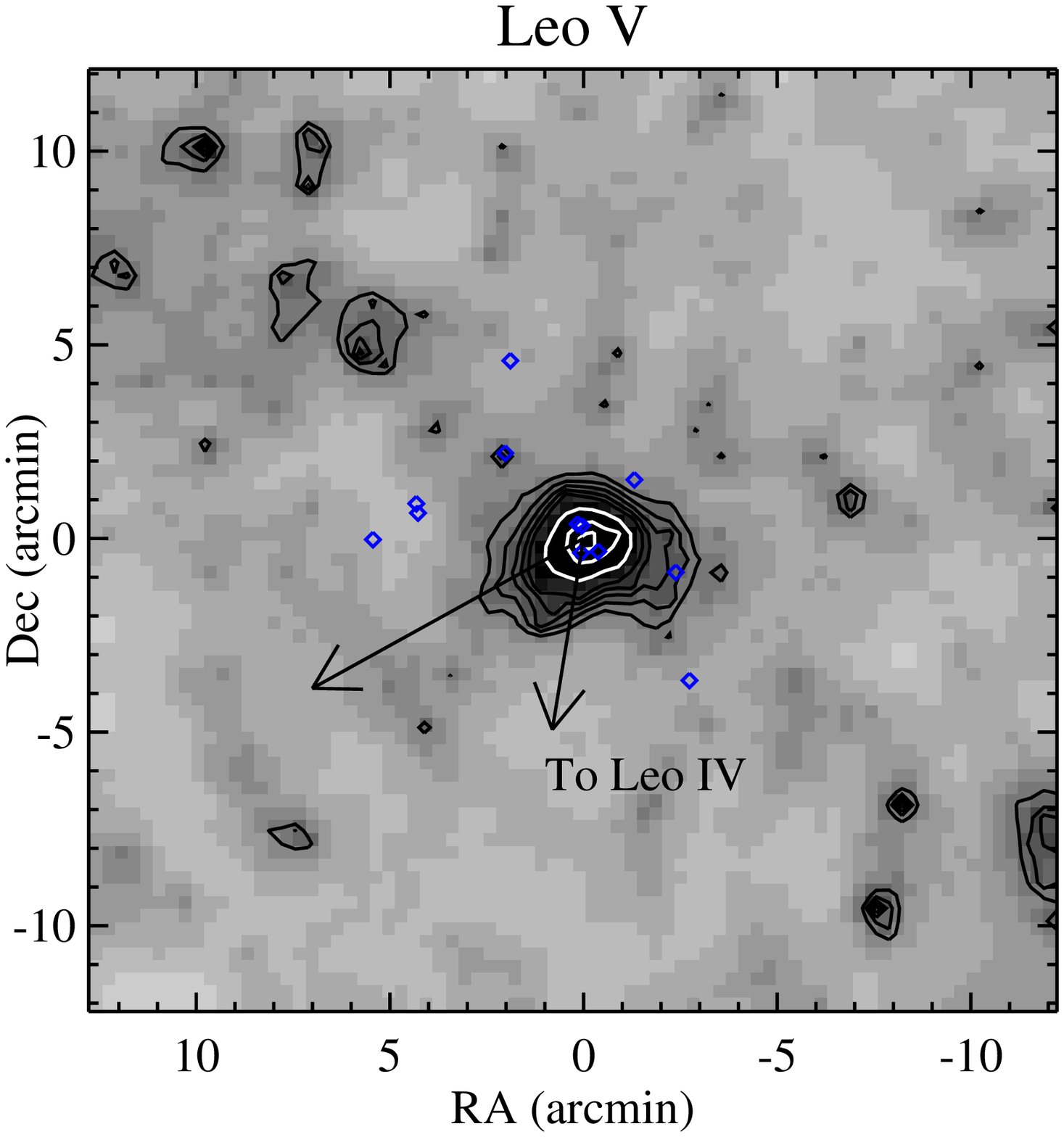} \epsfysize=6.0cm
\epsfbox{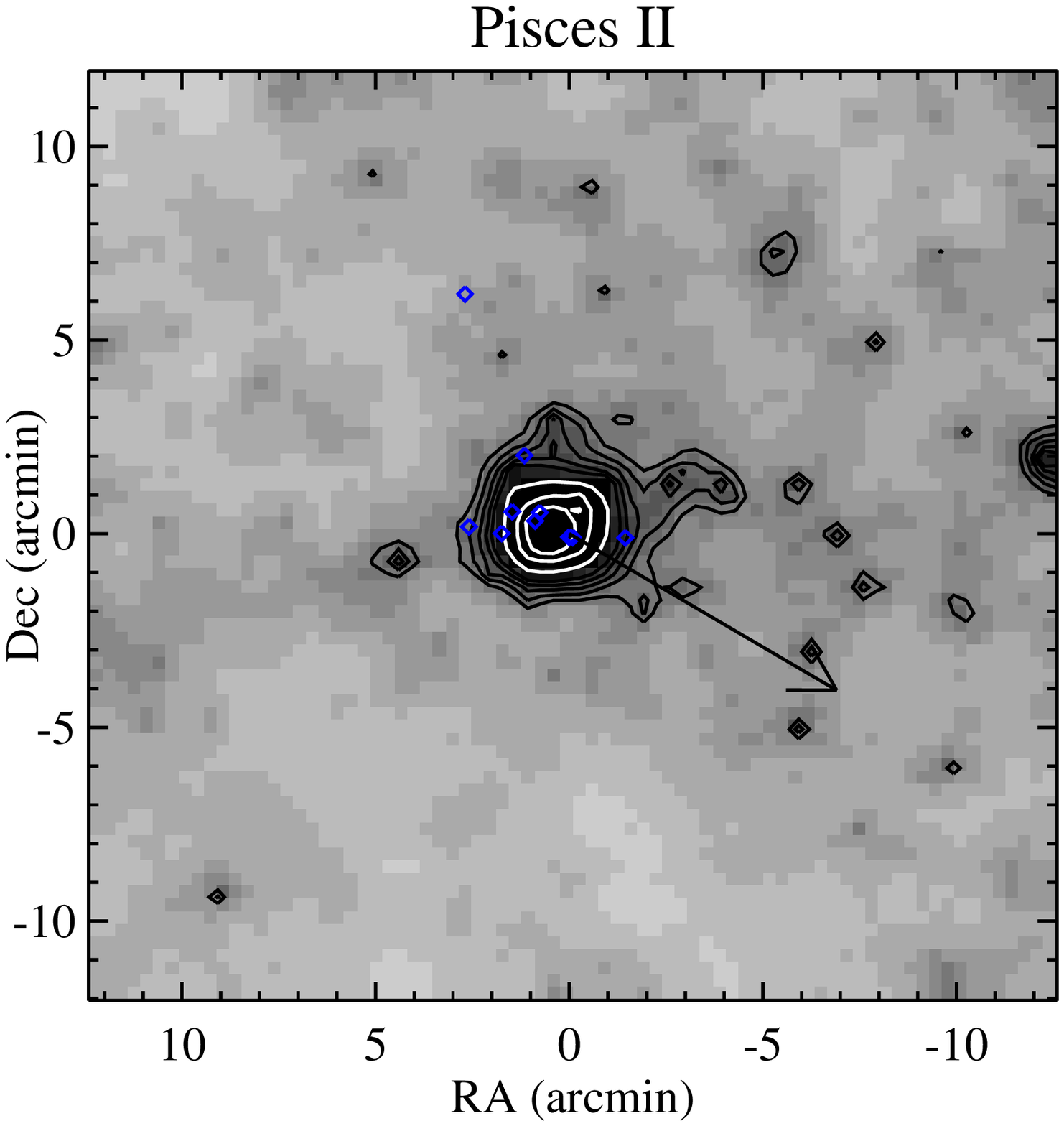} \epsfysize=7.0cm \epsfbox{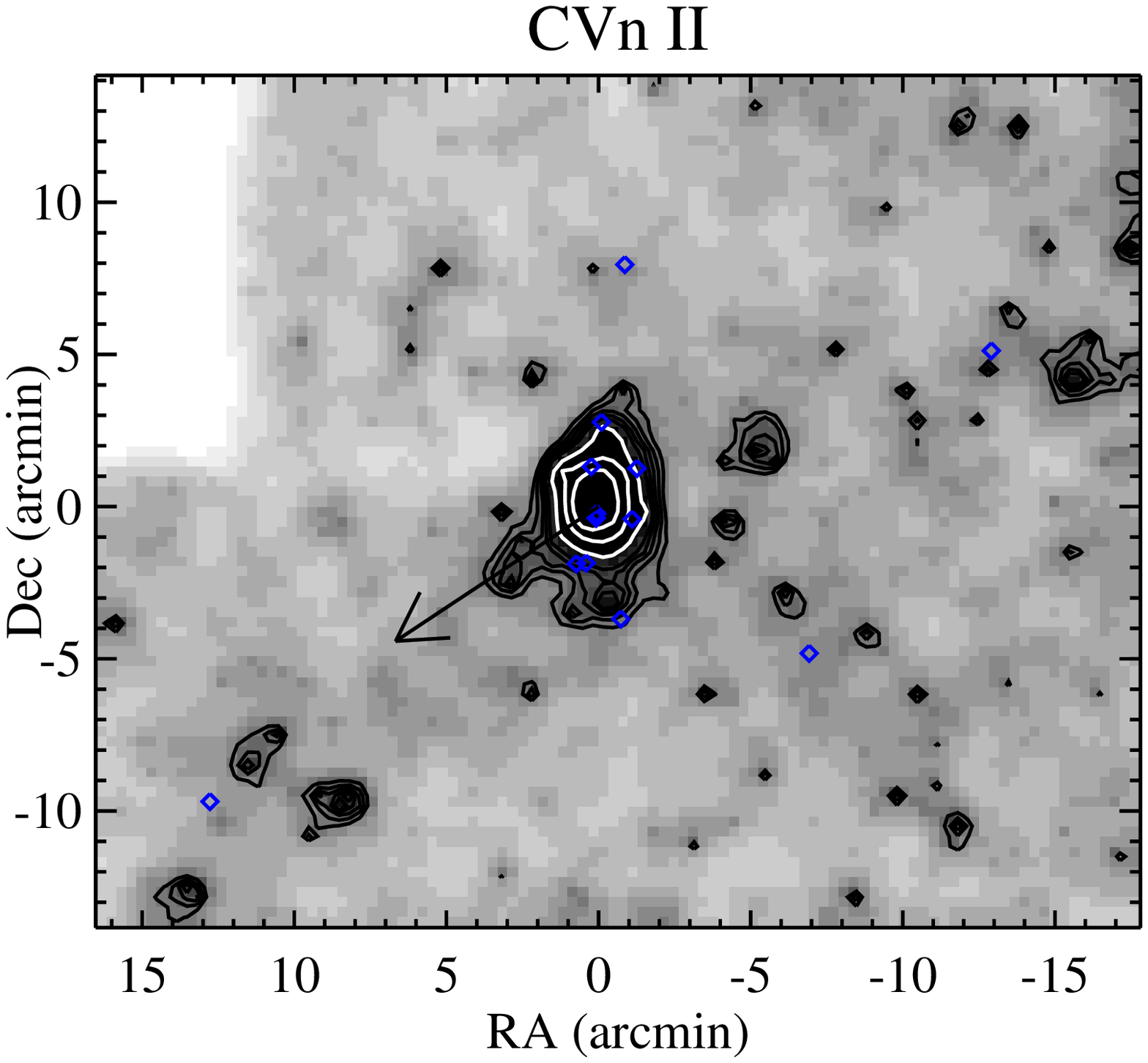}} 
\caption{Top row -- Matched filter maps of Leo~V, Pisces~II and CVn~II, with the contour levels showing the 5, 6, 7, 10, 15 and 20 $\sigma$ levels above the modal value in each map.  The pixel size in each map is 30 arcseconds, with no smoothing.  Each dwarf is clearly visible in the center of each field.  Bottom row -- Matched filter maps of Leo~V, Pisces~II and CVn~II, with the contour levels showing the 3, 4, 5, 6, 7, 10, 15 and 20 $\sigma$ levels above the modal value in each map (although, since these maps have been smoothed, these $\sigma$ levels cannot be thought of in the traditional sense).  The pixel size of each map is 20 arcseconds, and has been smoothed by a Gaussian with width 30 arcseconds.  The arrow in each plot points to the Galactic Center. The blue diamonds are likely BHB stars. Note that the second arrow in the Leo~V map is pointing in the direction of Leo~IV, with  no sign of extended structure in its direction.   Note the Subaru chip that was left out of our analysis for the CVn~II maps, as explained in \S~\ref{sec:datareduce}.
\label{fig:smoothmap}}
\end{center}
\end{figure*}

We show our final maps, both raw and spatially smoothed, for each of our satellites in Figure~\ref{fig:smoothmap}.  In the top row of Figure~\ref{fig:smoothmap} we show our raw matched filter maps with 30 arcsecond pixels, with contours showing (5, 6, 7, 10, 15, 20) $\sigma$ regions above the modal value of the map.  The main body of the satellite, along with several other overdensities, are visible in each map.  The bottom row of Figure~\ref{fig:smoothmap} show smoothed versions of the raw map.  For each of these we have binned our stars into 20 arcsecond pixels and then smoothed our final values using a Gaussian with width 1.5 times that of the pixel size.  The mode of the background of these smoothed maps was determined using the {\sc MMM} routine in IDL.  The contours shown in the plot are (3, 4, 5, 6, 7, 10, 15, 20) standard deviations above the modal value of the map, although their interpretation should be taken with a grain of salt, given that the maps were smoothed.  The vector arrow in each plot shows the direction to the Milky Way center.  We also show the vector to Leo~IV in our map of Leo~V; the two satellites are projected $\sim$2 degrees apart on the sky, although we see no sign of connection or interaction between the two (see \citet{Sandleoiv} and \citet{leoivleov} for further discussion).  We discuss possible satellite alignment with the MW in \S~\ref{sec:align}.  

There are tentative hints of extended structure in our satellite maps.  For instance, there are a series of overdensities going from the Northeast to Southwest in the map of Leo~V, and from Southeast to West in CVn~II, both of which apparently go through the body of the satellite.  We have constructed maps of barely resolved galaxies, by taking those objects culled due to their poor DAOPHOT PSF fits (see \S~\ref{sec:datareduce}), and only a couple of compact galaxy groupings are nearly coincident with those in the maps of our satellites in Figure~\ref{fig:smoothmap}.  Nonetheless, better star/galaxy separation, perhaps through near infrared photometry or Hubble Space Telescope imaging, will be necessary to definitively ferret out any possible streams in these systems.  The surface brightness at the $3-\sigma$ contour level is ($\mu_{g},\mu_{r}$)=(30.4,30.0) and (29.6,29.2) mag arcsec$^{-2}$ for  Pisces~II and Leo~V, respectively. CVn~II has surface brightness limits of ($\mu_{V},\mu_{I}$)=(30.3,29.6) mag arcsec$^{-2}$.

\subsubsection{The distribution of BHB stars}

We have also overplotted the spatial positions of BHB star candidates onto our maps in Figure~\ref{fig:smoothmap}, which are marked as blue diamonds.  Initial BHB star lists were taken from the blue box region in color-magnitude space in Figure~\ref{fig:CMDs} for each satellite, which were then further culled of objects which were not clearly on the BHB sequence.  This conservative approach may cause us to remove a time-varying RR Lyrae star candidate which truly does belong to a satellite, but does allow us to evaluate the spatial extent of the BHB population with the minimum amount of foreground/background contamination.  

As can be seen from Figure~\ref{fig:smoothmap}, the BHB star population is centrally concentrated around each satellite, although BHB stars at large radii are evident.  We take a simple approach to determine if our BHB spatial distributions are consistent with the parameterized model fits we derived for our satellites in \S~\ref{sec:paramfit}, which utilized stars on the main sequence to red giant branch ridgeline for parameter estimation.  We performed Monte Carlo simulations, randomly placing $N_{BHB}$ stars down at radii drawn from our Plummer and exponential models (varying the half light radius according to its uncertainty) and then comparing the median radius of the simulated sample of BHB stars  with the observed median.

According to our simulations, the BHB population of Pisces~II and CVn~II are consistent with the derived parameterized fit to the ridgeline data.  However, given our Leo~V structural parameters for the Plummer (exponential) profile fit, there is only a 0.4\% (0.3\%) chance of getting a BHB sample with an observed median radius that is at least as small as that measured for the bulk of Leo V stars -- strong evidence that the BHB stars in Leo~V are more extended than the fitted profile \citep[see also][]{leov}.  

Intrigued by the possibility that Leo~V may have an extended BHB stellar distribution, and seeking additional evidence, we ran our ML structural analysis code (see \S~\ref{sec:paramfit}) directly on our BHB sample.  The advantage of using the ML code is that it naturally incorporates a background surface density ($\Sigma_{b}$) into the structural measurement.  In this case, we found a half light radius of $r_{h}=2.9\pm0.8$ arcmin for our BHB stellar sample (assuming an exponential profile), which is $\sim$1.8$\sigma$ discrepant from that found for Leo~V's ridgeline stars.  While tantalizing, we can not say definitively that Leo~V has an extended BHB stellar distribution.

\section{Stellar Population and Luminosity}\label{sec:sfh}

In this section, we first perform a qualitative analysis to show which single population stellar ages and metallicities are consistent with our three satellite CMDs.  We then utilize the structural properties found in \S~\ref{sec:paramfit}, and our newly acquired knowledge of the satellites' stellar population to calculate their absolute magnitudes.

\subsection{General Properties} \label{sec:stellpop}

As can be seen from the CMDs within $2r_h$ in Figure~1, our data reach between $\sim$0.5 and 1 magnitudes below the main sequence turnoff.  Generally, in order to infer a precise age and metallicity of a stellar population, it is essential to have high quality photometry around and below this turnoff \citep{match}.  In our case, we are also fundamentally limited by the small number of satellite stars --  between $\sim$200-400, depending on the satellite -- as inferred from our parameterized structural analysis (\S~\ref{sec:paramfit}).  Because of these limitations, we eschew a CMD-fitting approach to inferring star formation history, and opt to qualitatively assess which single age stellar populations are consistent with our CMDs, given a range of metallicities that bracket the measured values in the literature.

\begin{figure*}
\begin{center}
\mbox{ \epsfysize=4.0cm \epsfbox{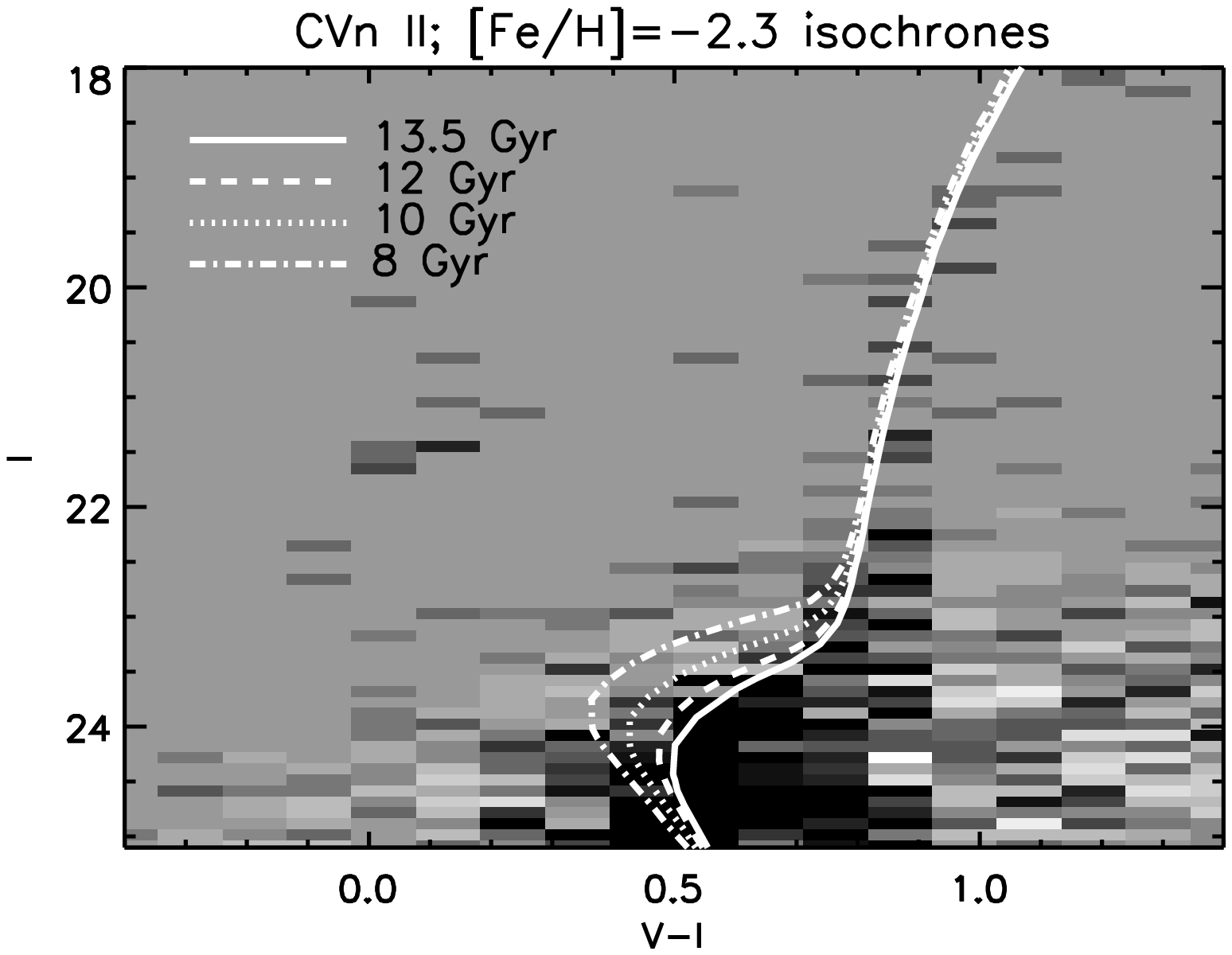} \epsfysize=4.0cm
\epsfbox{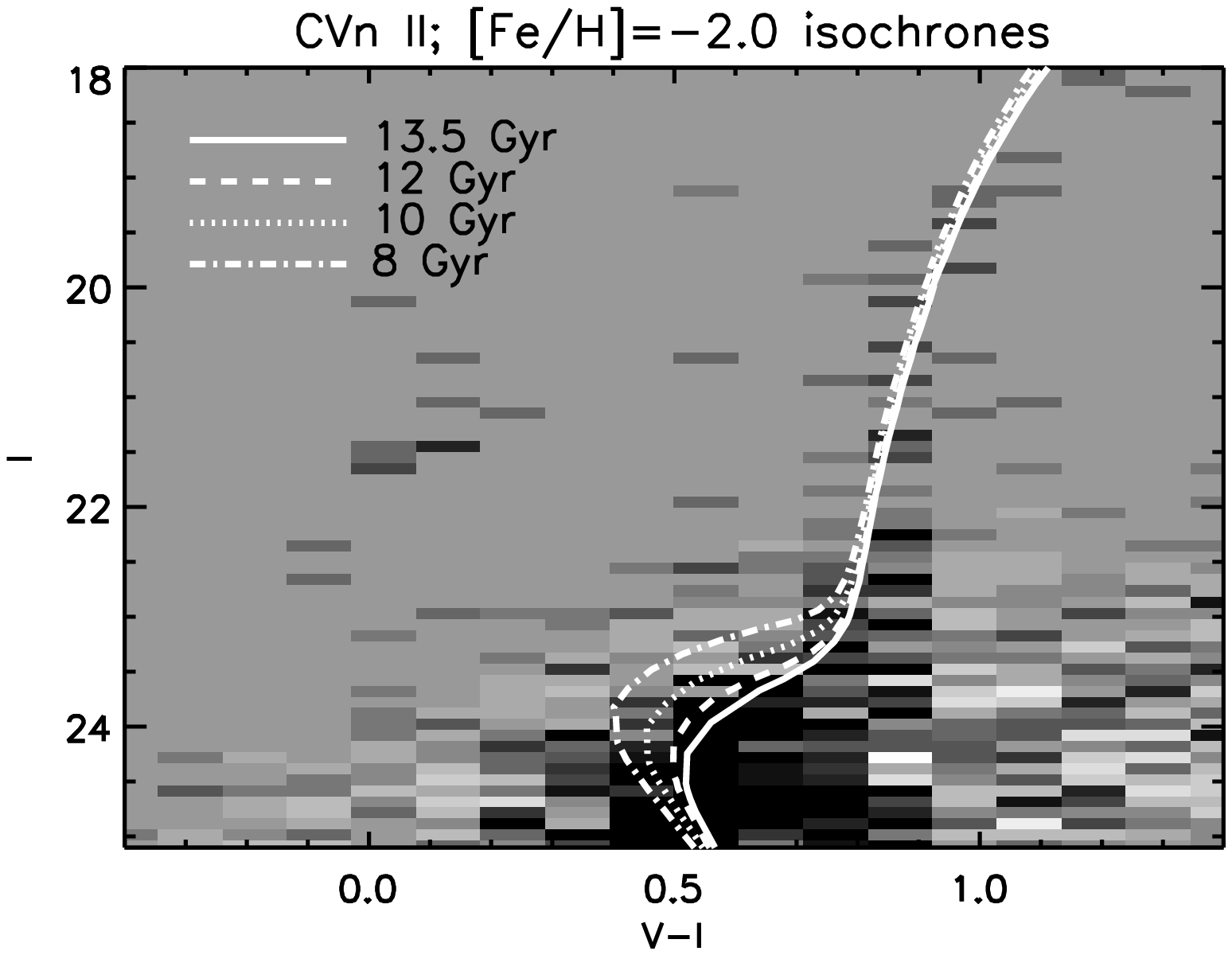}  \epsfysize=4.0cm
\epsfbox{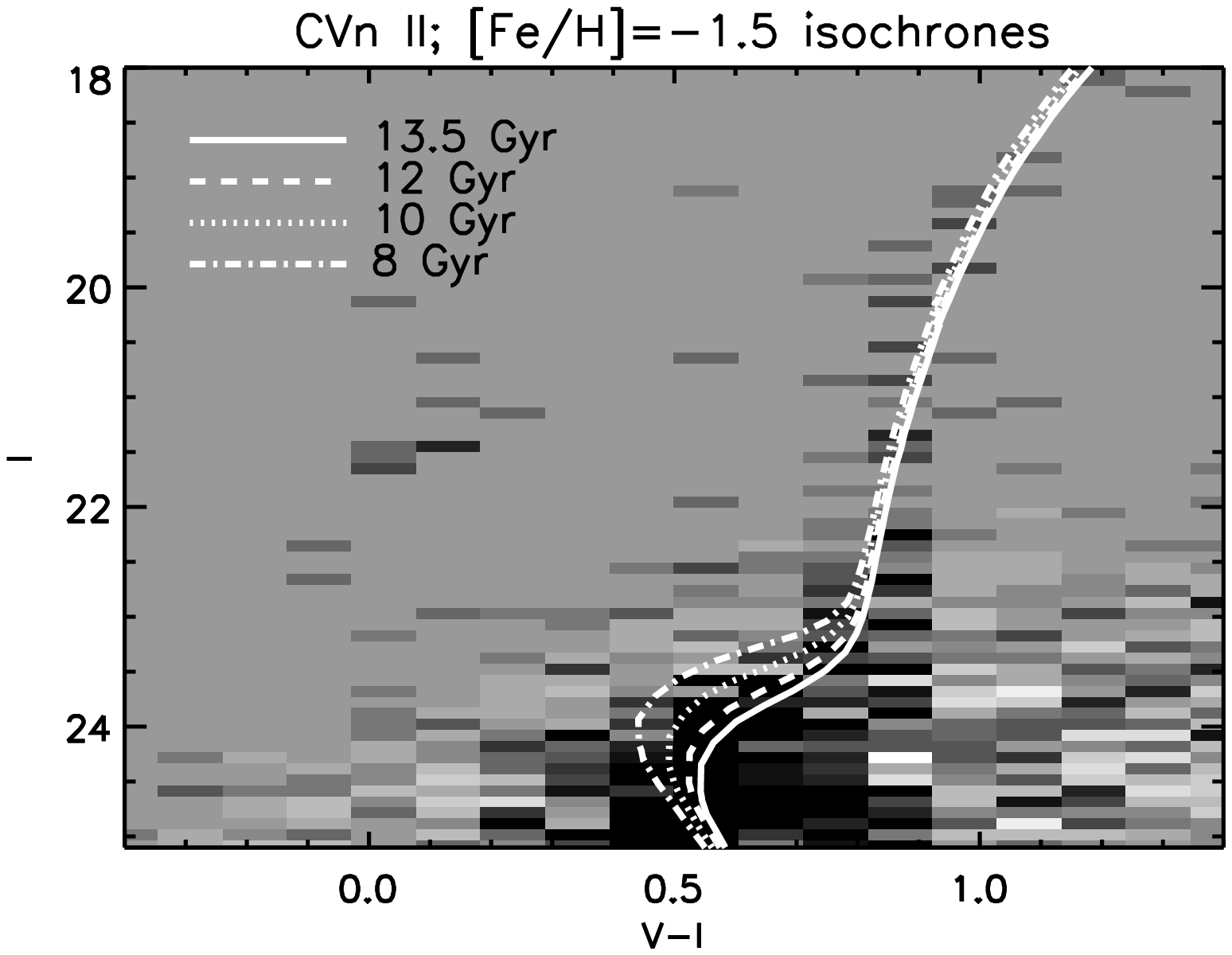}} 
\mbox{ \epsfysize=4.0cm \epsfbox{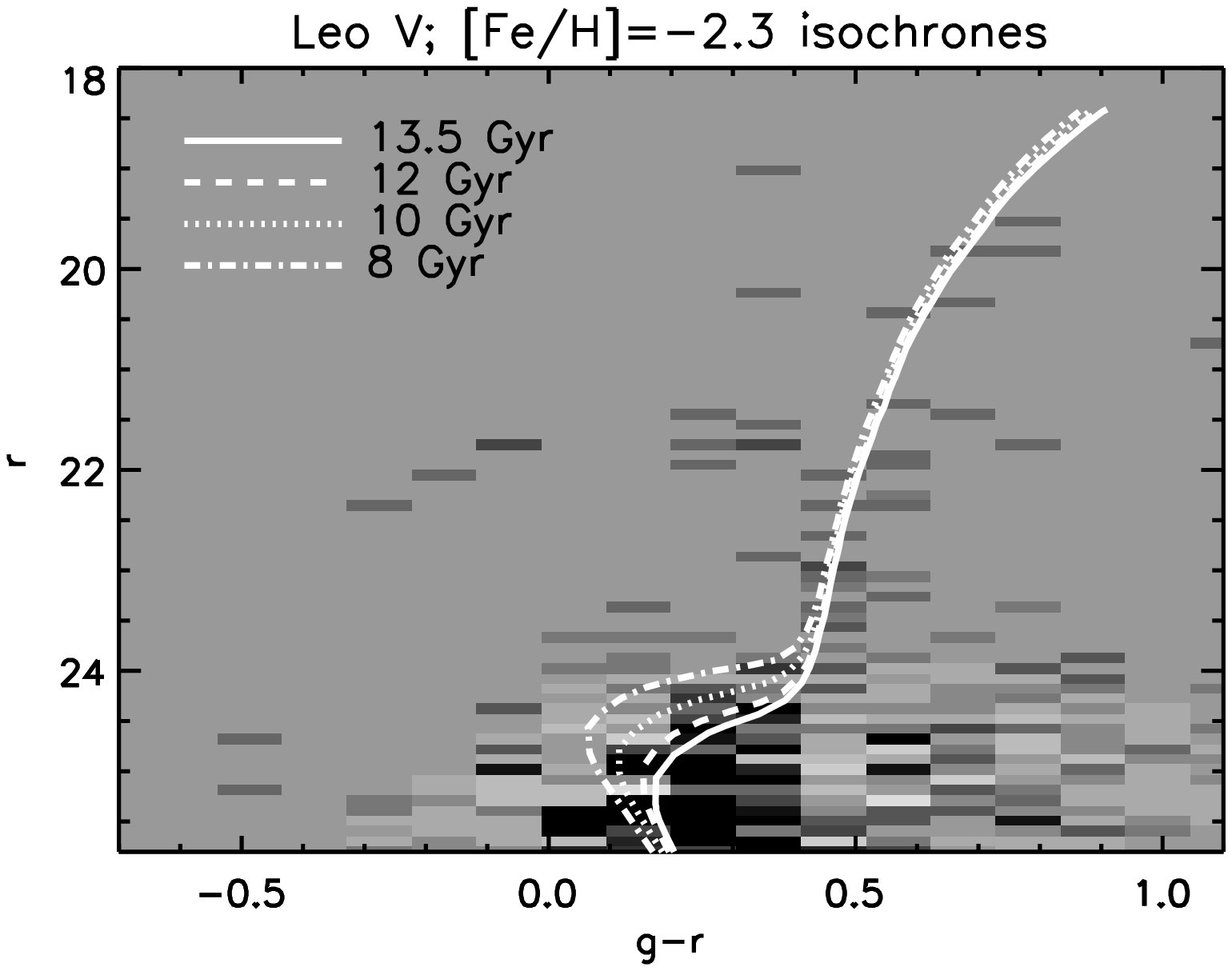} \epsfysize=4.0cm
\epsfbox{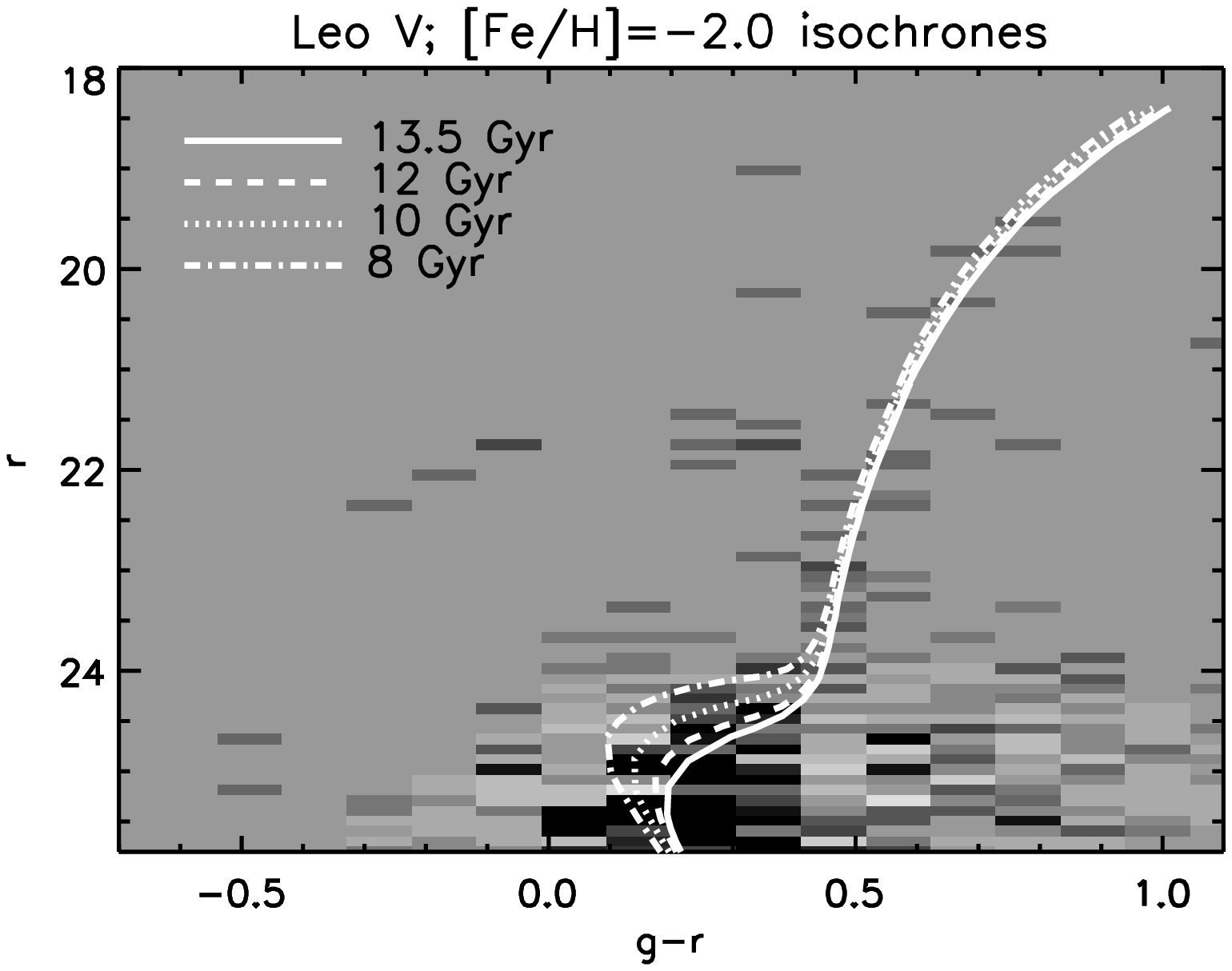}  \epsfysize=4.0cm
\epsfbox{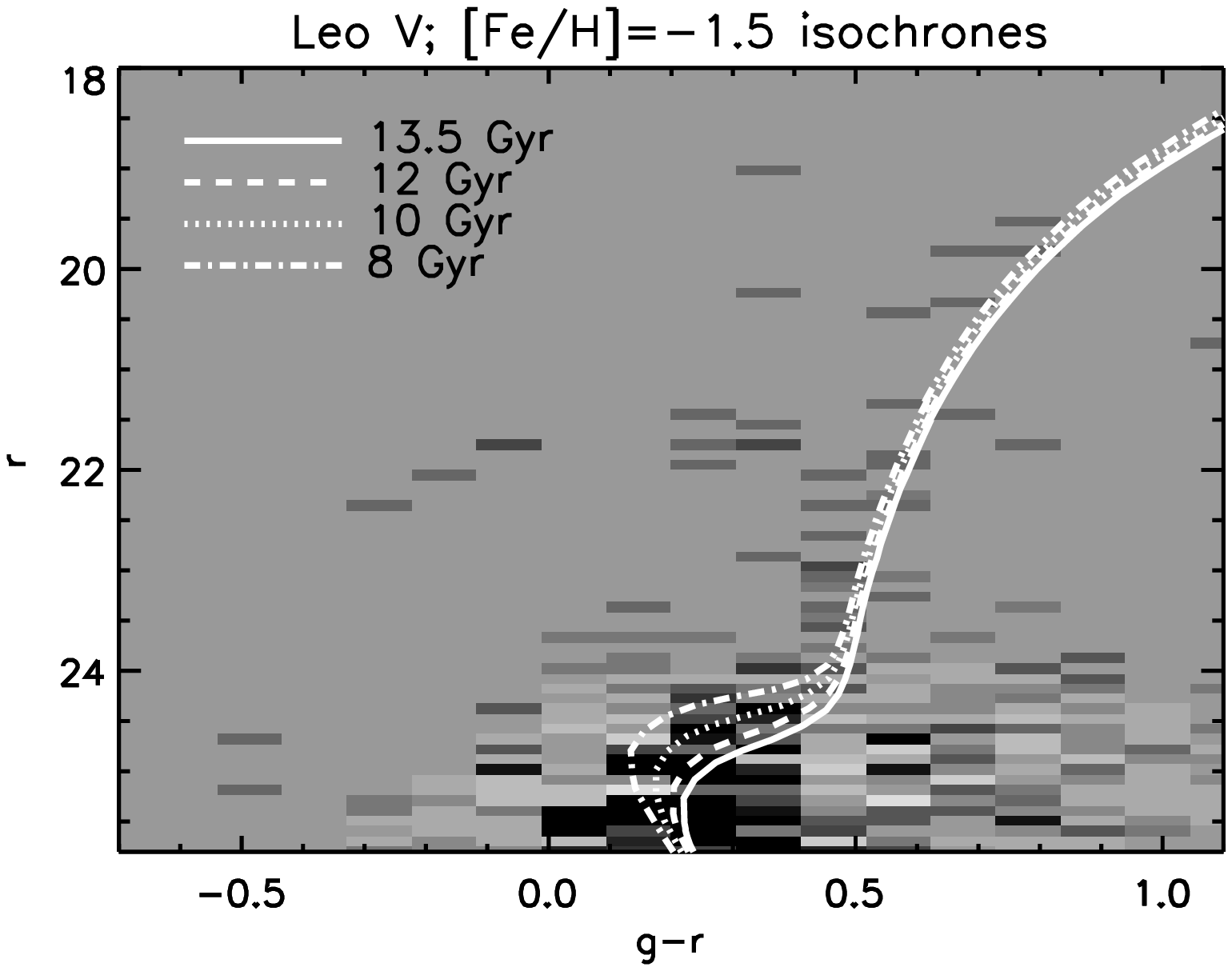}} 
\mbox{ \epsfysize=4.0cm \epsfbox{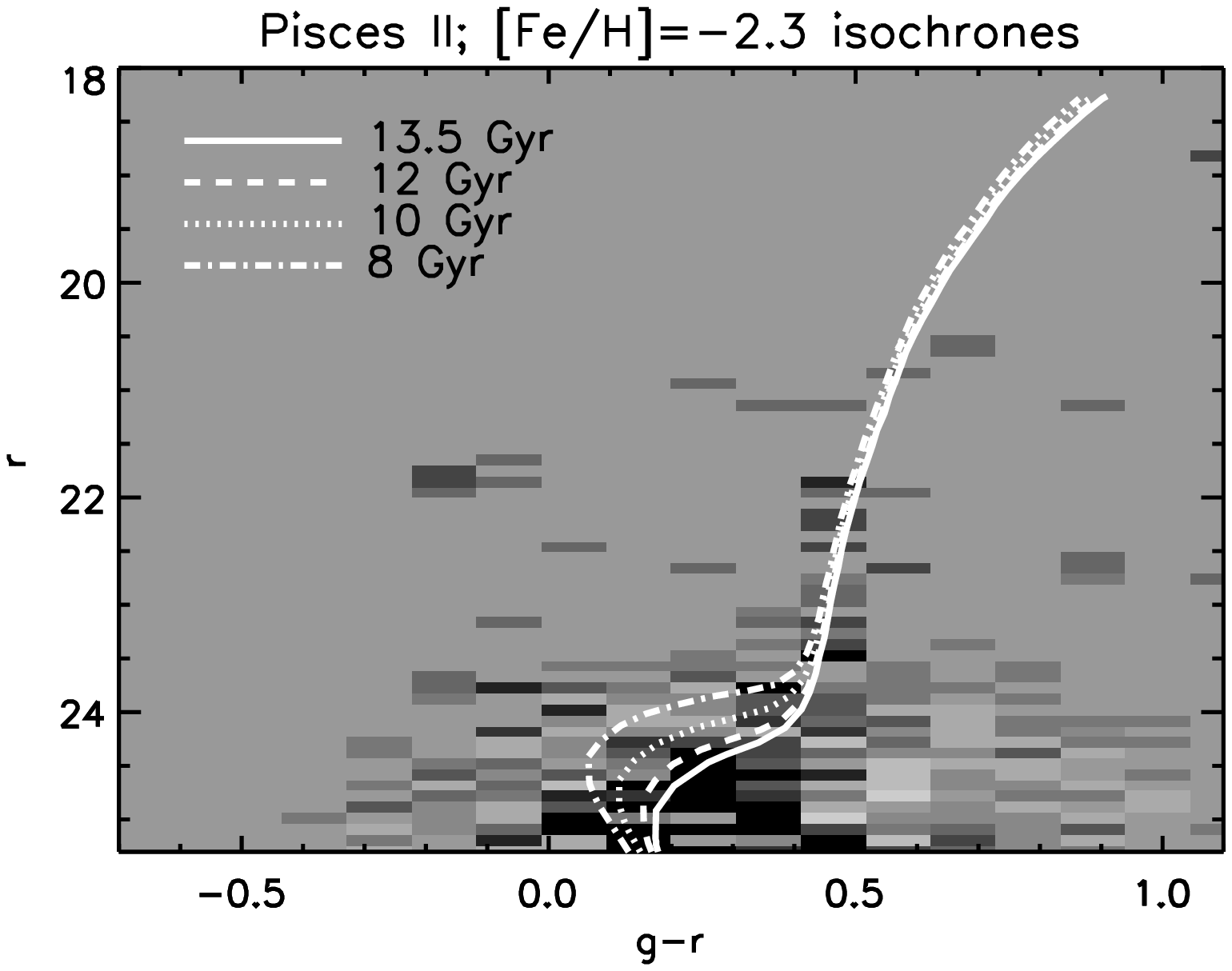} \epsfysize=4.0cm
\epsfbox{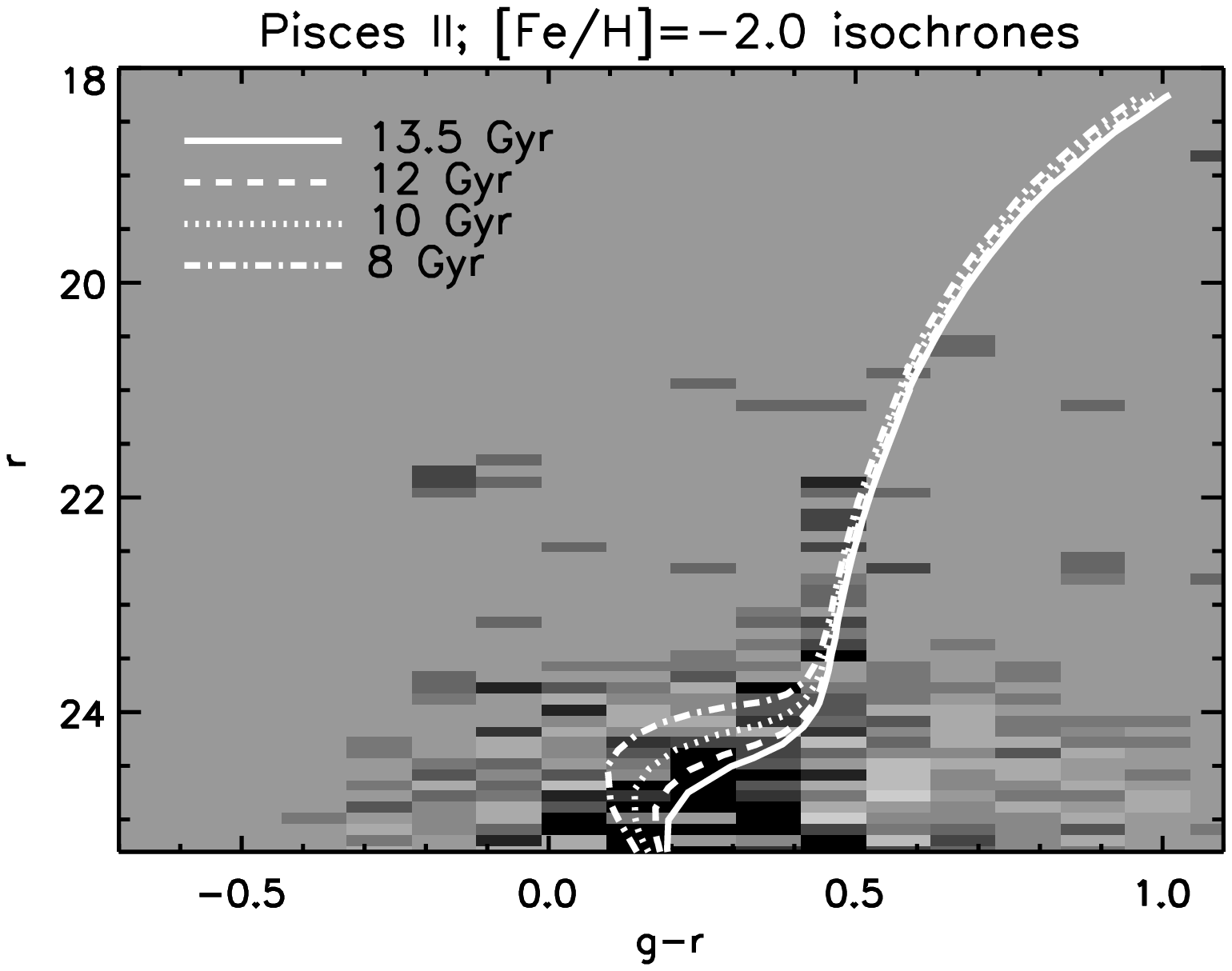}  \epsfysize=4.0cm
\epsfbox{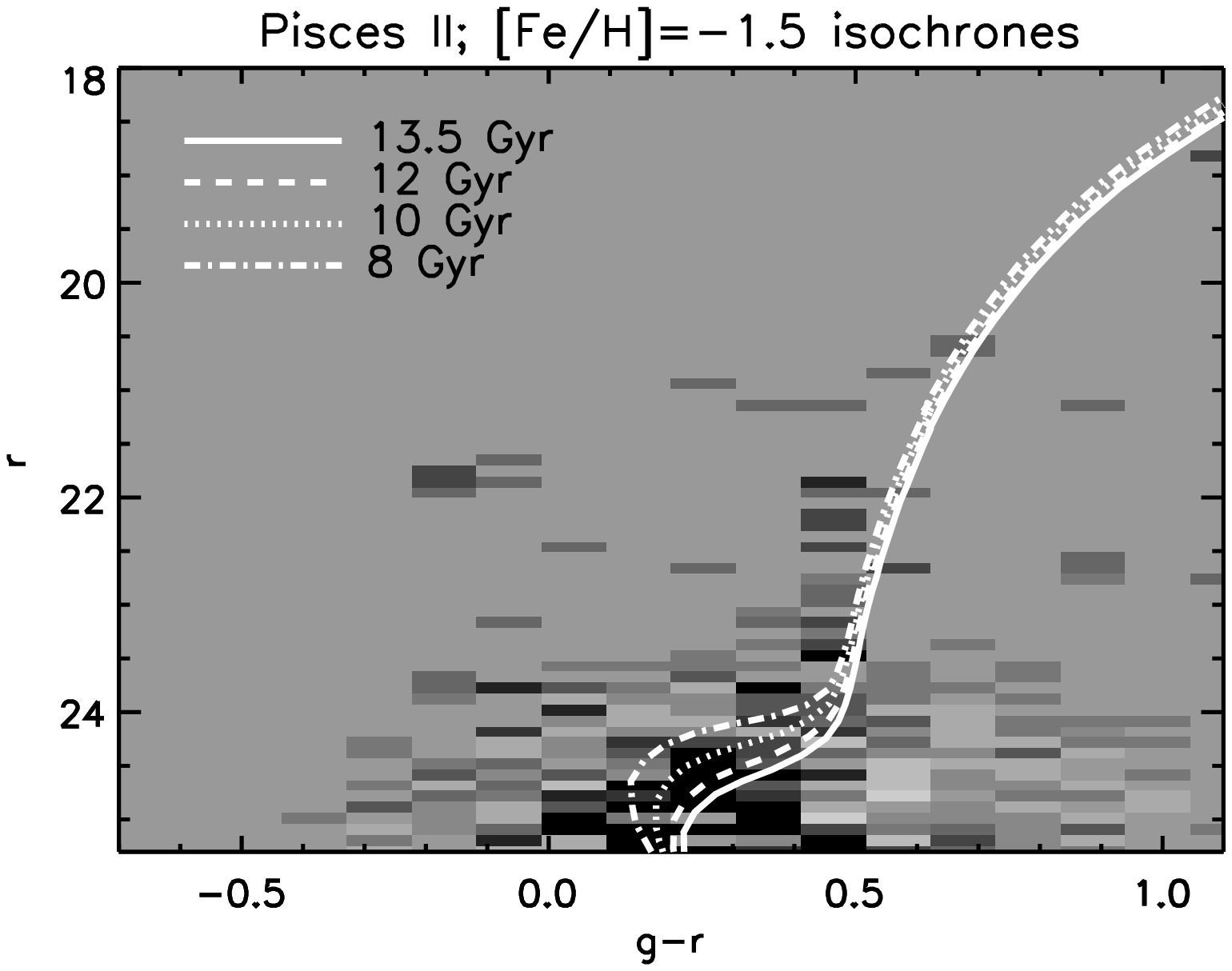}} 
\caption{Hess diagrams of CVnII, Leo V and Pisces II within 2$r_{h}$, with a background CMD subtracted.  We have also overplotted theoretical isochrones from \citet{Girardi04} in order to qualitatively assess the stellar populations which are compatible with the data -- see \S~\ref{sec:stellpop} for details. \label{fig:SFH} }
\end{center}
\end{figure*}

For each satellite, we plot a Hess diagram of all stars within $2r_h$, subtracting out a scaled background CMD based on stars greater than 8 arcminutes from the satellites' center (see Figure~\ref{fig:SFH}).  We overplot single age theoretical isochrones with age = (8,10,12,13.5) Gyr for a [Fe/H] of ($-2.3$,$-2.0$,$-1.5$), using the theoretical isochrone set of \citet{Girardi04} throughout.  The choice of metallicities brackets the mean spectroscopic metallicity measurements of CVnII \citep{Kirby08} and Leo~V \citep{Walker09}, although the dispersion to very low metallicities seen in CVn~II can not be modeled since the \citet{Girardi04} isochrone set has a minimum [Fe/H] of $-2.3$.  We also assume that Pisces~II has a comparable metallicity as the other two satellites, a relatively safe assumption since all of the new MW satellites studied thus far have spectroscopic $-2.6 < $ [Fe/H] $< -2.0$ \citep{Kirby08}.

In addition to the theoretical isochrone set of \citet{Girardi04}, we have compared our CMDs against the Dartmouth Stellar Evolution Database \citep{Dotter07,Dotter08}. These two theoretical isochrone sets, at least for old and metal poor stellar populations, have been shown to be consistent by \citet{Dotter07}, although the red giant branch in the isochrones of \citet{Girardi04} are slightly hotter (bluer).  For the purposes of our qualitative assessment of the age and metallicity of Leo~V, CVn~II and Pisces~II, we have verified that our conclusions are not dependent on the particular theoretical isochrone set.  These isochrone sets are also consistent with the CMDs of old, metal-poor globular cluster colors in $g$ and $r$ \citep{Dotter07,An08}.  \citet{Willman11} showed that metal-poor red giant branch stars observed in Draco and Willman 1 match the colors predicted by Dotter isochrones in $g-r$ (but not in other SDSS colors).

The plots in Figure~\ref{fig:SFH} simply illustrate the stellar age range compatible with each satellite for a given metallicity.  In summary, all three of our objects appear to be old ($>10 $ Gyr) and metal poor ([Fe/H]$\sim-$2).  Even stellar ages of 10 Gyr appear to be only marginally consistent with our CMDs, and only then at metallicities slightly higher than spectroscopic measurements indicate.  In this sense, the three objects in the current study are consistent with nearly all of the newly discovered faint MW satellites that have been studied in detail \citep[e.g.][]{sdsssfh,Martin08,sandherc,Sandleoiv,Okamoto12}.  

Leo~V, Pisces~II and CVn~II do not have a clear overabundance of blue plume stars as has been reported for CVn~I \citep{Martin08} and Leo~IV \citep{Sandleoiv}, two satellites at similar Galactocentric distances ($D>150$ kpc).  The overabundant blue plume stars in CVn~I and Leo~IV have been interpreted as evidence for a recent ($\sim$2 Gyr) star formation episode (although see also Okamoto et al. 2011).  Therefore, to the best of current observational limitations, Leo~V, Pisces~II and CVn~II have SFHs that appear qualitatively different than the other distant MW dwarfs.  We will discuss the observational limitations and possible implications in \S~\ref{sec:context}.

\subsection{Absolute Magnitude}

As has been pointed out previously, measuring the total magnitude of the new MW satellites is problematic due to their very sparse stellar content \citep[e.g.][]{sdssstruct}.  The luminosity of one of these faint satellites can change significantly simply by the addition or removal of a few red giant branch stars.  To account for this `CMD shot noise', we mimic the luminosity measurements of previous work \citep[e.g.][]{sdssstruct,sandherc,Sandleoiv}, which we briefly describe here.

First we created several realistic, well-populated model CMDs from the \citet{Girardi04} theoretical isochrone set (assuming a Salpeter initial mass function) using the {\it testpop} program within the StarFISH software suite.  See \S~\ref{sec:extend} for the procedure for generating well-populated model CMDs to represent each satellite.  For Leo~V and Pisces~II, we used a set of four isochrones which correspond to their observed stellar populations (see \S~\ref{sec:stellpop}): 1) [Fe/H]=$-$2.3 with an age of 13.5 Gyr; 2) [Fe/H]=$-$2.0 with an age of 13.5 Gyr; 3) [Fe/H]=$-$2.3 with an age of 12.0 Gyr; and 4) [Fe/H]=$-$2.0 with an age of 12.0 Gyr.  For CVn~II, we use model isochrone sets with [Fe/H]=$-2.0$ and [Fe/H]=$-1.5$, each with 12.0 and 13.5 Gyr stellar populations, since these better correspond to our findings in \S~\ref{sec:stellpop}.

We drew 1000 random realizations of the model CMDs, each with the same number of stars we found  in our parameterized structural analysis for each satellite (\S~\ref{sec:paramfit}), and determined the `observed' magnitude of each realization above our 90\% completeness limit.
A correction was made for stars below our completeness limit by using luminosity function corrections derived from \citet{Girardi04}, assuming a Salpeter initial mass function.  For a given model CMD, we take the median value of our 1000 random realizations as the measure of the absolute magnitude of the satellite and its standard deviation as the uncertainty, presuming it had a stellar population with the same age and metallicity as the model CMD.  We also varied the presumed distance to each satellite by its 1-$\sigma$ uncertainty and recomputed its absolute magnitude, using the offset from the best value as an estimate of the uncertainty in absolute magnitude associated with the uncertainty in distance modulus.  This distance modulus related uncertainty was added in quadrature with the CMD shot noise uncertainty described above, although it is a subdominant factor.   To convert from $M_{r}, M_{g}$ magnitudes to $M_{V}$ magnitudes (when necessary) we use the filter transformation equations of \citet{Jordi06}.

The value reported in Table~\ref{table:paramfits} is the average absolute magnitude found for  the four single population isochrones used for each satellite, where the uncertainty includes both the typical uncertainty on each individual measurement and the spread among the four isochrones, added in quadrature.  

Our derived absolute magnitude for each satellite is consistent, at the $1-\sigma$ level, with the latest measurements in the literature, albeit with smaller uncertainties, with one exception.  Using nearly identical Subaru data to our own, \citet{Okamoto12} found $M_{V}=-5.37\pm0.2$ mag for CVn~II, significantly brighter than our own measurement of $M_{V}=-4.6\pm0.2$.  Although a difficult measurement, the origin of this discrepancy is unclear, and is puzzling given the agreement in the other structural properties we derived.

\section{Discussion}\label{sec:discuss}

\subsection{Leo~V, Pisces~II and CVn~II in Context}\label{sec:context}

We have presented deep imaging of three of the recently discovered MW satellites -- Leo~V, Pisces~II and CVn~II -- all of which are faint ($M_V > -5$) and distant ($150 \lesssim D \lesssim 200$ kpc).  At these distances, these objects are nearly the least luminous detectable in the SDSS survey given their size, surface brightness and luminosity \citep[e.g.][]{Walsh09,Koposov08}.  Utilizing wide field imaging (with extent $\sim$10 times that of the satellites' half light radius) and photometry deeper than the main sequence turn off, we studied the star formation history and precisely constrain the structural parameters of these objects.  We also performed a careful search for signs of disturbance or disruption which would not be picked up by our parameterized fits, but might yield clues about their origin.

All three satellites have similar stellar populations; they are old ($>10$ Gyr) and metal poor ([Fe/H]$\sim-2$). This is the norm, with little variation, among the post-SDSS MW satellites, excepting the distant, transitional galaxy, Leo~T \citep{Irwin07,leot}. Given the paucity of stars in each object, not to mention the intrinsic limits of current theoretical stellar isochrones \citep[and their inability to gauge absolute ages; see][for a discussion]{Marin09}, it was impossible for us to truly distinguish between 12 Gyr and 13.5 Gyr stellar populations (for instance) and we therefore cannot comment on whether these objects are candidate ``reionization fossils" \citep[e.g.][]{Ricotti05,Gnedin06,Bovill11a}, a proposed population of MW satellites whose star formation occurred primarily before the epoch of reionization.

We see no evidence in our satellites for the extended star formation that has been observed in all four classical dSphs more distant then $\sim$90 kpc \citep[see][for a review]{Dolphin05} and in the two other ultra-faint dwarfs at $D \gtrsim$ 150 kpc, CVn~I \citep{Martin08} and Leo~IV \citep{Sandleoiv}.  If Leo~V, Pisces~II, and CVn~II lack any intermediate or young ($<$10 Gyr) stellar population (although we can not rule this out at the moment; see next paragraph), then environment does not solely determine a satellite's SFH.  It may be that the details of infall and orbital history play a role \citep[e.g.][]{Mateo08,Rocha11}, or that some minimum initial baryonic or dark matter reservoir may be necessary for extended SF.  Such a finding would paint a more nuanced picture of galaxy formation than provided by the classical dSphs alone.  The classical dSphs' observed correlation between SFH and MW distance has previously suggested the possible dominance of environmental processes in the truncation of star formation \citep[e.g.][]{vandenbergh94}.  For example, more nearby dwarfs have on average experienced more extensive ram pressure stripping of gas, local ionizing radiation, and pericentric passages, which could work in concert to exhaust their gas early on \citep{HZ04,ZH04}.  

We must cautiously interpret the lack of observed recent star formation in our satellites.  CVn~I is $\sim$25 times more luminous and Leo~IV is $\sim$2.5 times more luminous than the three satellites considered here.  If the stellar populations of CVn~I and Leo~IV were scaled down by a factor of 25 and 2.5, respectively, then a small handful of blue plume stars, evidence for recent star formation, could remain.  However, it isn't clear whether such scaled down versions of CVn~I and Leo~IV would be outliers from the blue straggler/BHB star relation (N$_{\rm BSS}$/N$_{\rm BHB}$, \citealt{Momany07}) that has been used as evidence for young stellar populations in these objects.  Moreover, if Leo~V, Pisces~II, or CVn~II had undergone a small burst of star formation at intermediate ages (e.g. 5-8 Gyr), it may not have been resolved by the present study given the small numbers of stars in these dwarfs and given the lack of theoretical isochrones at [Fe/H] $<-2.5$.  We note that none of the ultra-faint satellites with $D < 150$ kpc, except perhaps UMa~II \citep{sdsssfh}, show evidence for extended star formation.

CVn~II and Leo~V show direct hints of stellar mass loss through the stream-like structures tentatively observed in our matched filter maps, and more tenuous indirect hints through their [Fe/H] and through Leo~V's alignment with the Galactic center.  Leo~V has a spectroscopic [Fe/H] of $\sim-2.0$ \citep{leovspec}, nearly 0.5 dex higher than expected for its luminosity of $M_{V}=-4.4$ mag, and CVn~II has a spectroscopic [Fe/H] of $-2.2$, $\sim$0.2 dex higher than expected based on the positive correlation between dwarf luminosity and [Fe/H] observed by \citet{Kirby11}.  While at face value, this may indicate that at one time Leo~V and CVn~II were more luminous, neither are strong outliers from the [Fe/H] vs. luminosity relation of \citet{Kirby11}.  As can be seen in Figure~\ref{fig:smoothmap}, Leo~V's major axis is roughly aligned with the direction to the MW, which might be another hint of its disturbed structural state (see \S~\ref{sec:align} for further discussion).  

Wide field kinematics and deep imaging with improved star/galaxy separation (e.g. with HST or near-infrared photometry), will provide a clearer picture of Leo~V's and CVn~II's evolutionary states.  The present evidence appears strong enough to add Leo~V to the list of the new MW satellites -- including Ursa Major I \citep{Okamoto08}, Ursa Major II \citep{Munoz10}, Hercules \citep{Coleman07,sandherc,Aden09} and Willman 1 \citep{Willman06,Willman11} -- which show clear signs of disturbance or extended structure in their structural and/or kinematic properties.

\subsection{Structural Properties of the New Satellites}\label{sec:structall}

At this point, several new MW satellites have been discovered since the work of \citet{sdssstruct}, who did a comprehensive overview of the structure of the ultra-faint satellites based on the SDSS discovery data.  Additionally, several works have shown that data as shallow as the SDSS, while sufficient to discover the new satellites, can only crudely constrain their size \citep{sandherc,Munoz11}.  Thus, we seek to revisit and extend this initial analysis, and have compiled a table of salient properties of both the post-SDSS satellites and the classical dSphs in Table~\ref{table:structparam}. To compile this list, we took the latest literature values, emphasizing those structural results derived via a maximum-likelihood method similar to that of \citet{sdssstruct} when possible.  We do not include Segue~3 \citep{Belokurov10,Fadely11}, Koposov~1 or Koposov~2 \citep{Koposov07} into the table, as these objects have properties most similar to globular clusters.  We also do not include Leo~T, which is a distant, transitional galaxy \citep{Irwin07,leot}.

\begin{figure*}
\begin{center}
\mbox{ \epsfysize=5.5cm \epsfbox{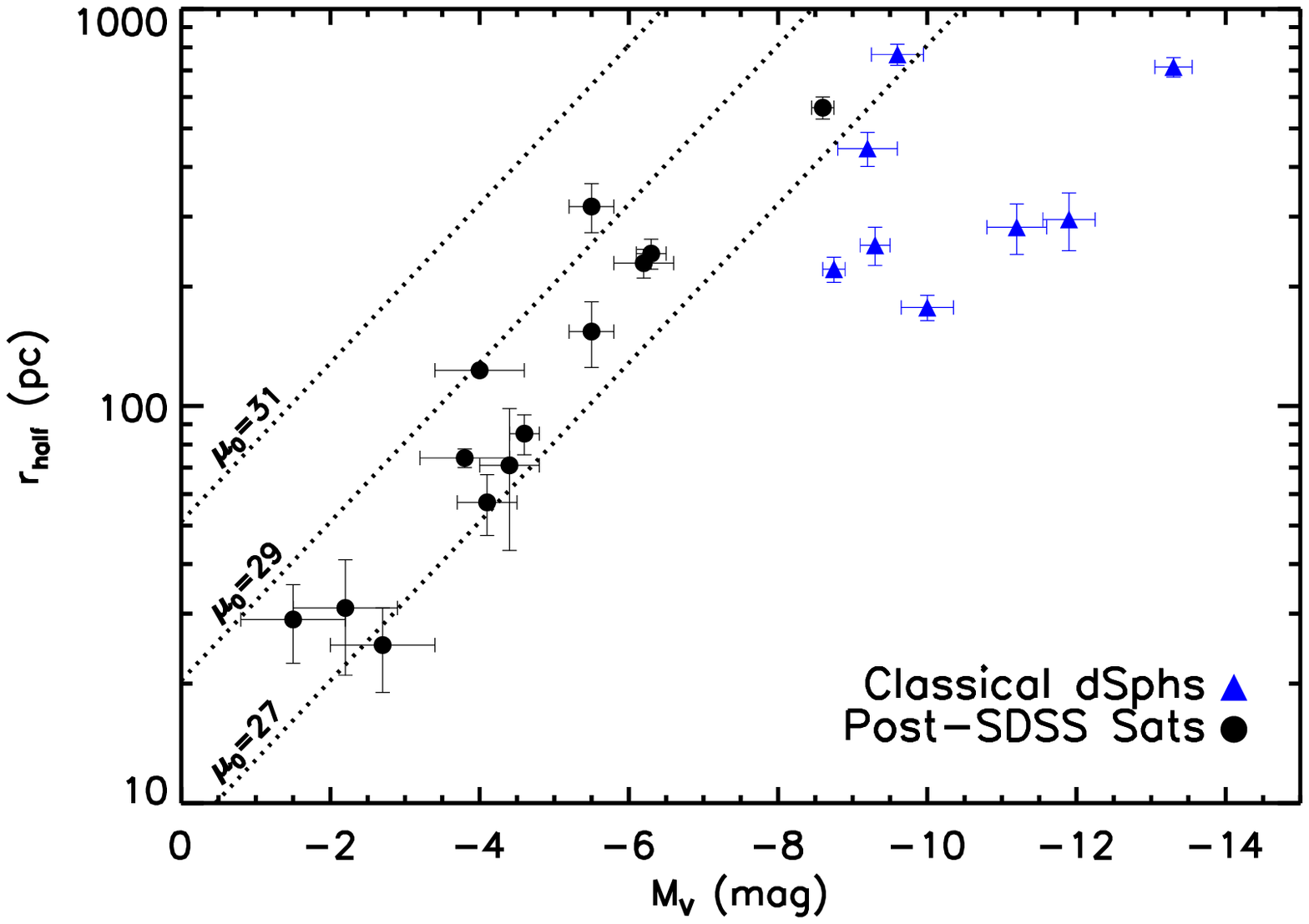} \epsfysize=5.5cm
\epsfbox{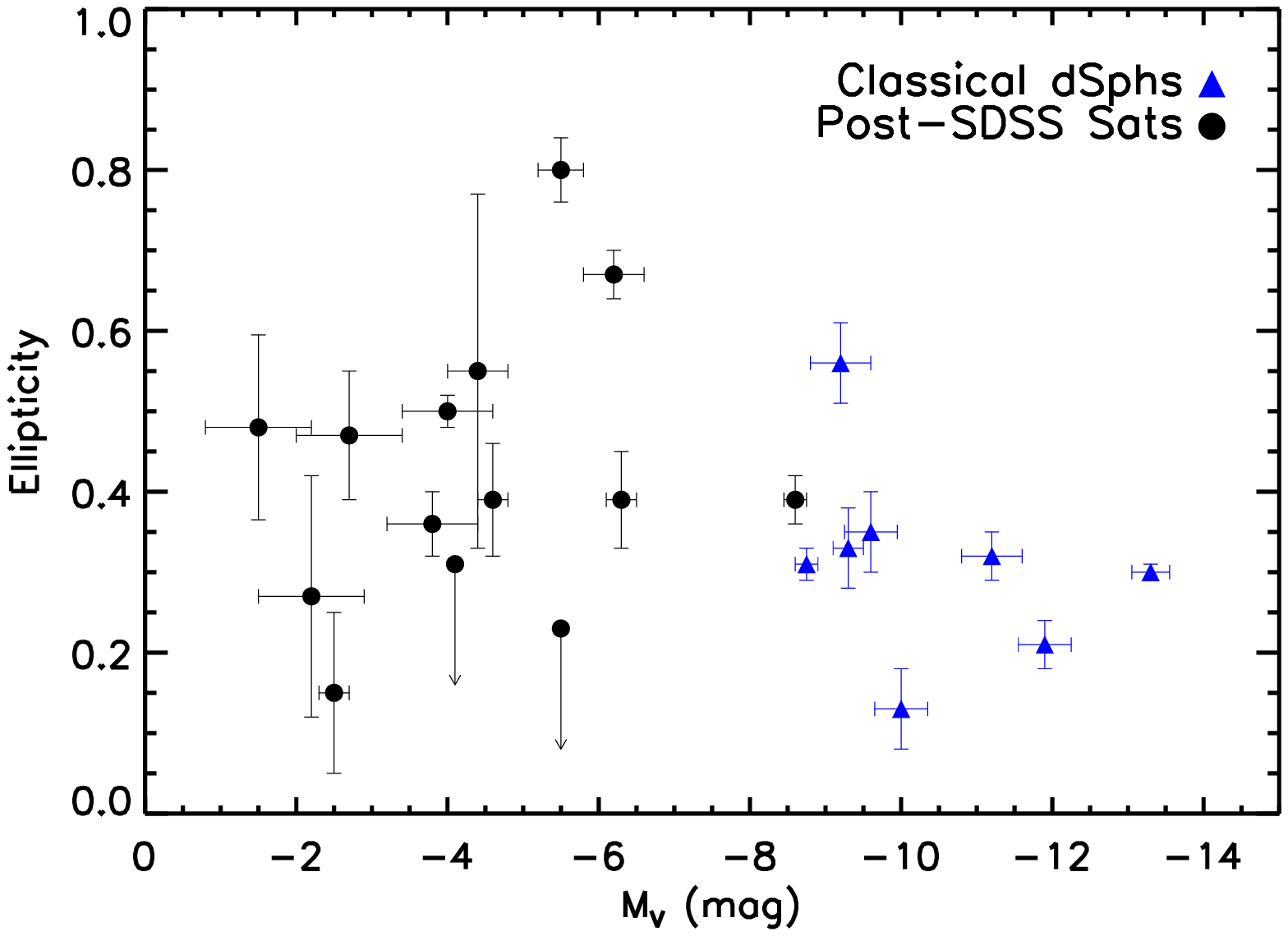}} 
\caption{Distribution of the MW satellites in the $M_{V}$ vs. $r_{half}$ (left) and $M_{V}$ vs. $\epsilon$ (right) planes.  Note that CVn~I, although discovered in the SDSS, has similar properties as the classical dSphs in both panels.  The apparent decline in half-light radius as a function of magnitude among the new MW satellites may be artificial, as objects that are `large and faint' would have gone undetected by the SDSS survey \citep{Walsh09,Koposov09}, but would sit in the upper left corner of the plot.  Lines of constant surface brightness are shown to highlight this selection effect.  As discussed in \S~\ref{sec:structall}, there is little statistical evidence that the post-SDSS satellites and the classical dSphs have a different ellipticity distribution, as portrayed in the right panel.  However, Hercules ($\epsilon=0.67$) and Ursa Major I ($\epsilon=0.80$) still stand out in this plot.  \label{fig:MVrelations} }
\end{center}
\end{figure*}

We begin by summarizing the whole data set in the $M_{V}$ vs. $r_{half}$ and $M_{V}$ vs. $\epsilon$ planes in Figure~\ref{fig:MVrelations}.  We note that CVn~I, discovered in the SDSS \citep{Zucker06cvni}, has similar structural properties as the classical dwarf spheroidals, albeit on the faint end of their distribution.  For the rest of this discussion, we will present results lumping CVn~I with the classical dSphs although it is marked as a `post-SDSS satellite' in the plots.  Despite the appearance of a monotonic rise in half light radius as a function of magnitude among the new satellites, there are certainly selection effects in this plot, because the SDSS-based searches were not sensitive to the `faint and large' objects that would populate the upper left corner of the plot \citep{Walsh09,Koposov09}, and which are predicted by the latest theoretical work \citep[e.g.][]{Bullock10,Bovill11b}.  It remains to be seen whether this trend is purely an artifact of selection effects, or reflects a true property of the faintest dwarf galaxies.

In the right panel of Figure~\ref{fig:MVrelations} we show the distribution of ellipticities as a function of luminosity for the MW satellites.  \citet{sdssstruct} argued that the new SDSS satellites have a tendency towards higher ellipticities than their classical counterparts, which we revisit.  Since Leo~IV \citep{Sandleoiv} and Pisces~II only have upper limits for their measured ellipticities we used  $\epsilon = 0.15\pm0.14$ and $\epsilon=0.18\pm0.17$ in the discussion that follows, respectively, since this range corresponds to the 16th and 84th percentile in our bootstrap analysis for each object.  A simple Kolmogorov-Smirnoff (KS) test yields an 87\% chance that the classical dSphs and the new satellites are not drawn from the same parent ellipticity distribution, which is not significant, and so we conclude that they may have been drawn from the same distribution.  Since the uncertainties on each individual data point are substantial, a KS test may be inappropriate.  Assuming the ellipticity distributions can be represented by a Gaussian, we use a maximum likelihood estimator to determine the mean ellipticity and dispersion of the classical dSph and new satellite ellipticity distributions given the different uncertainties on each individual measurements \citep[see][]{Pryor93}.  The new satellites have a mean ellipticity of $\langle \epsilon \rangle=0.44\pm0.05$, with a spread of $\sigma_{\epsilon}=0.17\pm0.04$, while the classical dSphs have $\langle \epsilon \rangle=0.32\pm0.03$ and a spread of $\sigma_{\epsilon}=0.09\pm0.03$.  The offset in mean ellipticity between the two data sets is only significant at 2.1$\sigma$, while the difference in spreads is only $1.6\sigma$ apart.  Thus, both a KS test and a maximum likelihood estimate of the ellipticity distributions indicate at most only a slight  difference between the new satellites and the classical dSphs.  Even if the ellipticity distribution of the new MW satellites is not statistically different from the classical dSphs, there are several new satellites with extreme ellipticities that deserve focused follow up to establish their nature -- for instance, Ursa Major I ($\epsilon=0.80$) and Hercules ($\epsilon=0.67$).

\subsection{Milky Way -- Satellite Orientation}\label{sec:align}

Intrigued by the near-alignment of the major axis of Leo~V with the direction to the MW center (Figure~\ref{fig:smoothmap}), along with a similar configuration in Coma Berenices \citep{Munoz10}, we investigated the orientation of all of the new MW satellites and classical dSphs.  One might naively expect, for instance, that if the new MW satellites (or some subpopulation of them) were being tidally disrupted by the MW, they would be preferentially oriented in the direction of the MW center on the sky (modulo projection effects, which tend to place any feature along the line of sight, given our relative proximity to the MW center). `Inner' tidal tails tend to form in the direction of the gravity vector while outer tidal tails (presumably with surface brightness too low to detect) would be oriented nearly in the direction of the satellite's orbit \citep[e.g.][]{Combes99,Dehnen04}.  Recently, \citet{Klimentowski09} confirmed this idea by simulating a dwarf galaxy (consisting of both baryonic and DM components) as it progressed on an eccentric orbit around a MW-like galaxy potential.  By taking snapshots of the dwarf as it orbited the MW and noting its orientation with respect to the MW center throughout, \citet{Klimentowski09} built up a histogram of angles between the MW center and the `inner' tidal tail of the dwarf galaxy.  While the dwarf galaxy went through the whole range of possible orientations, there was a broad peak in the angular offset between the dwarf and the MW center between $\sim10^{\circ} - 30^{\circ}$ \citep[see Figure 3 of][]{Klimentowski09}.

We compare the predictions of \citet{Klimentowski09} with the observed orientations of the MW satellites, relative to the Galactic center, to look for imprints of tidal effects on them as a populations.  To do this, we first assume that the observed position angles of MW satellites roughly correspond to the inner tidal tails described by \citet{Klimentowski09}.  We have measured the angle between a satellite's major axis and the direction to the MW center, $\Delta \theta_{GC}$, and present the results in Table~\ref{table:structparam}, along with the other structural properties of the new and classical dSphs.  We remind the reader that $\Delta \theta_{GC}$ could not be calculated for Leo~IV and Pisces~II, as their low apparent ellipticity results in an undefined major axis position angle.  As a population (where CVn~I is again considered to be more similar to the classical dSphs), the new MW satellites have a median $\Delta \theta_{GC}=30^{\circ}$.

\begin{figure*}
\begin{center}
\mbox{ \epsfysize=5.5cm \epsfbox{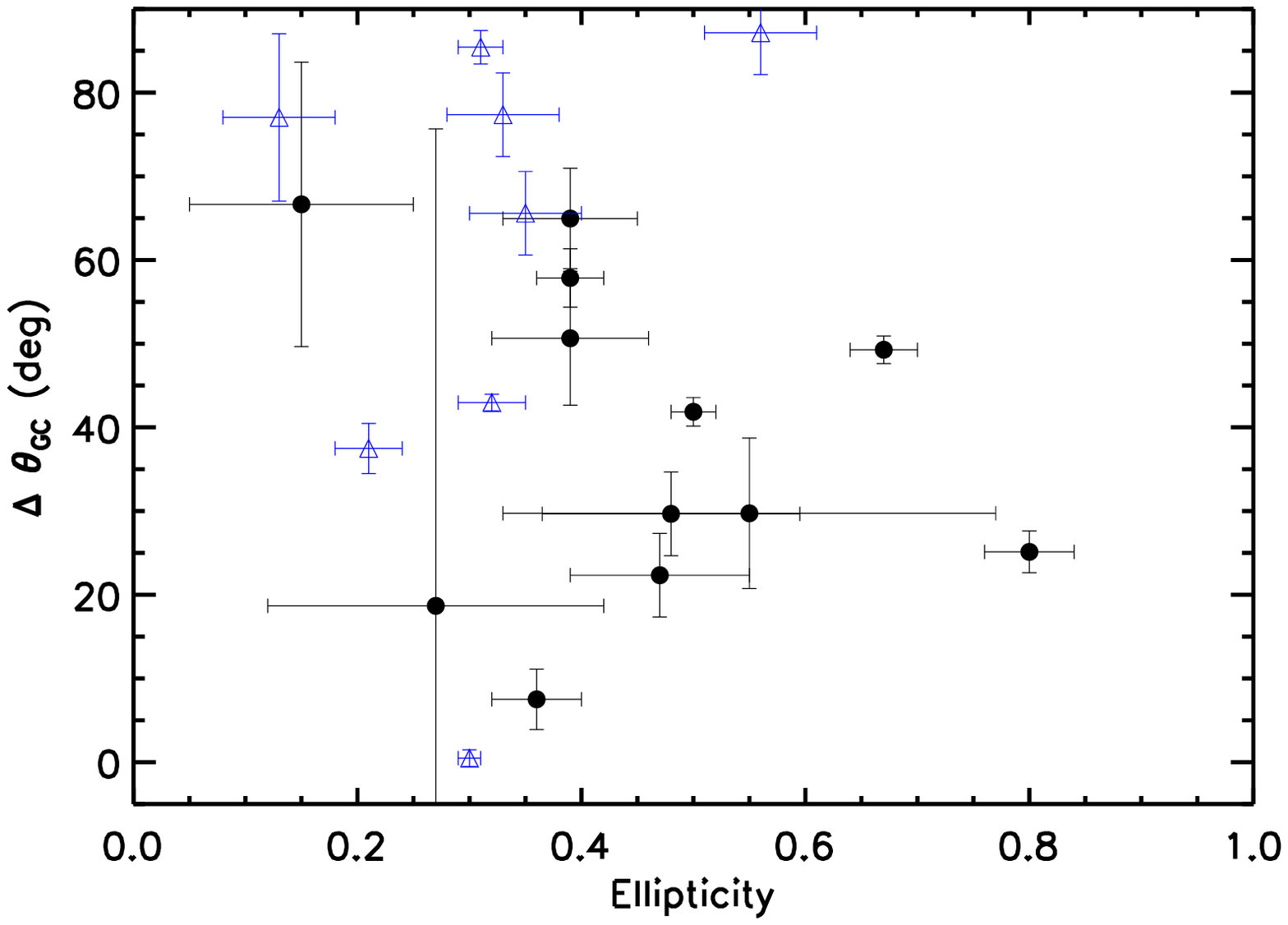} \epsfysize=5.5cm
\epsfbox{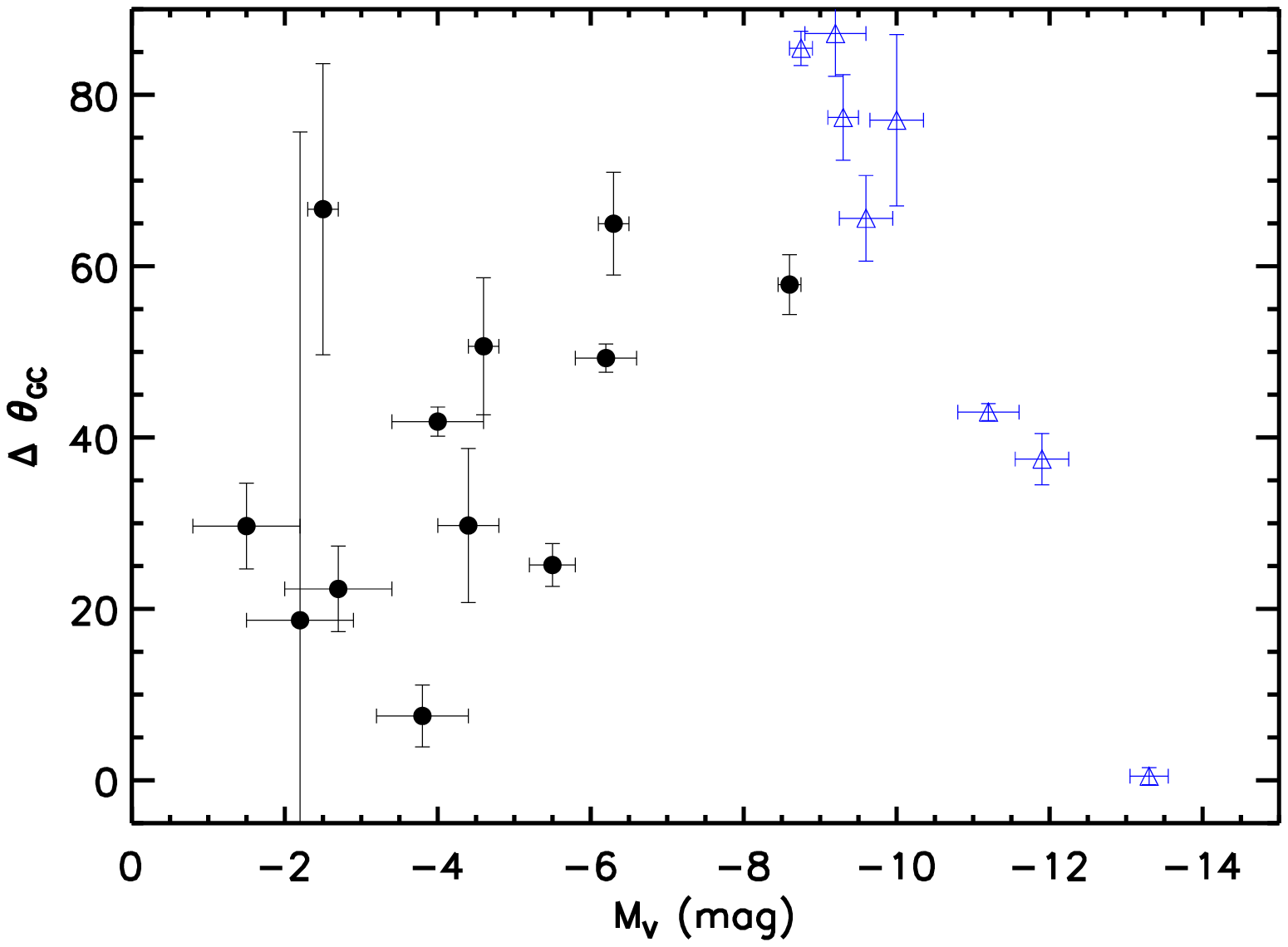}} 
\mbox{ \epsfysize=5.5cm \epsfbox{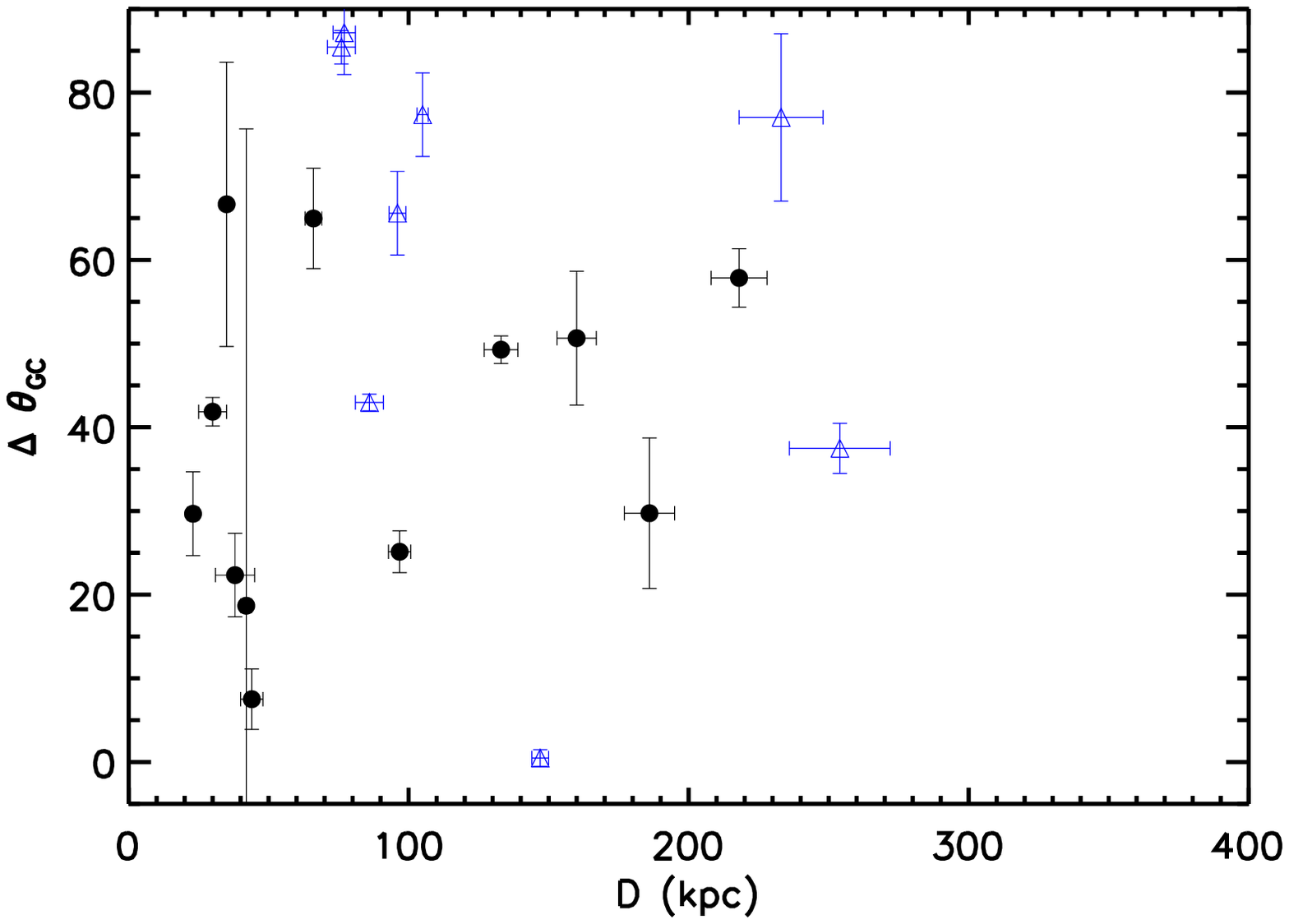} \epsfysize=5.5cm
\epsfbox{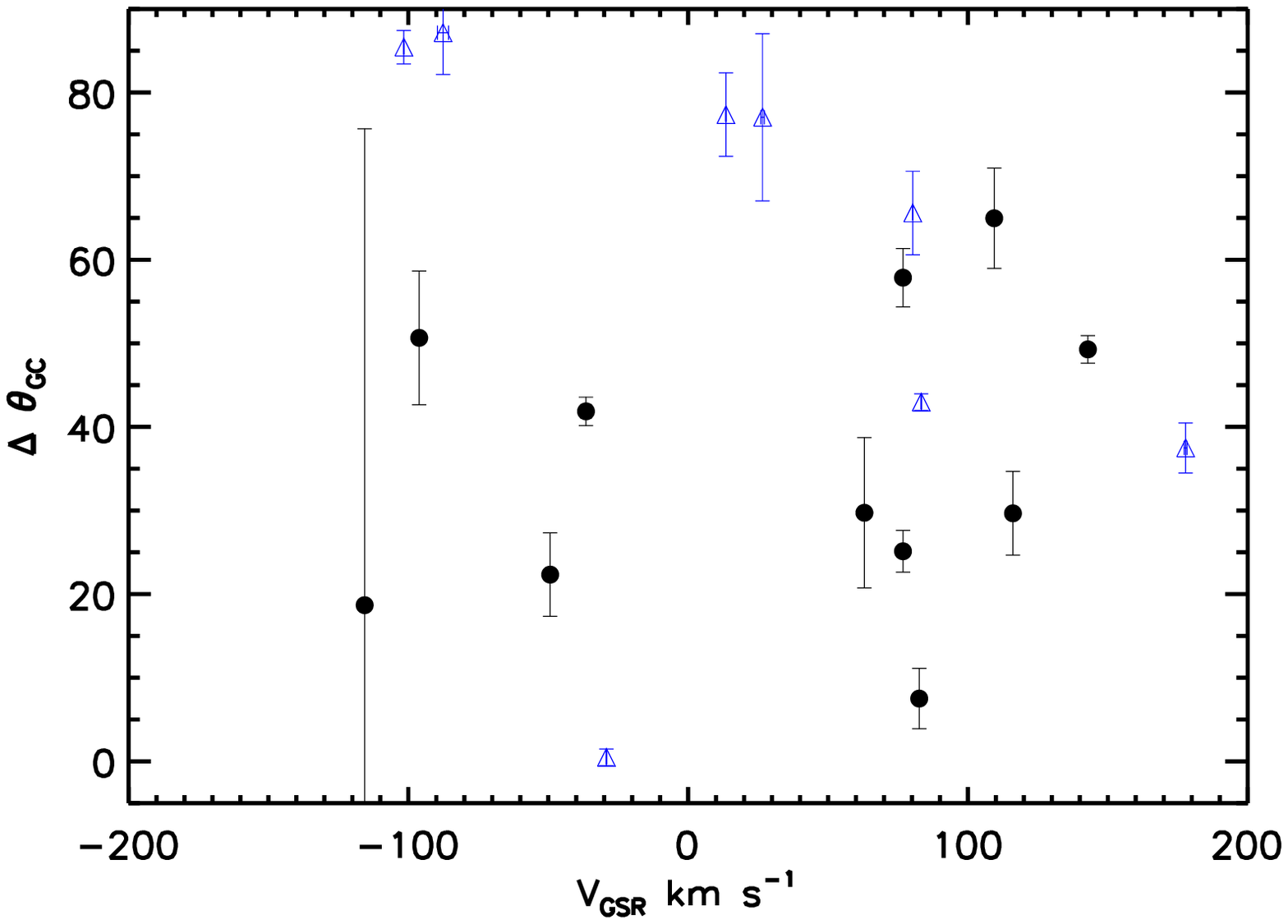}} 

\caption{Here we plot the difference in angle between the major axis of the MW satellites and the direction to the MW center ($\Delta \theta_{GC}$) as a function of various satellite properties.  As in Figure~\ref{fig:MVrelations}, the black circles represent the new MW satellites, while the blue triangles are the classical dSphs.  In the upper left panel, we plot $\Delta \theta_{GC}$ as a function of satellite ellipticity, $\epsilon$.  Note that the new MW satellites with the highest ellipticity ($\epsilon \gtrsim 0.4$) tend to have $20^{\circ} \lesssim \Delta \theta_{GC} \lesssim 40^{\circ}$.  In the upper right panel we show $\Delta \theta_{GC}$ as a function of satellite luminosity ($M_{V}$) and note that the faintest of the new MW satellites also are the most aligned with the MW center.  One obvious exception is the satellite Segue~2.  There is no apparent relation between distance to the MW, $D$, and $\Delta \theta_{GC}$; nor with the satellites' velocity with respect to the Galactic standard of rest ($V_{GSR}$), as can be seen in the bottom two panels.  Note that neither Pisces~II or Leo~IV are represented in these plots, as their position angle is not well determined.  \label{fig:thetaGC} }
\end{center}
\end{figure*}

In Figure~\ref{fig:thetaGC}, we have plotted $\Delta \theta_{GC}$ versus ellipticity ($\epsilon$), luminosity ($M_{V}$), distance and velocity with respect to the Galactic Standard of Rest ($V_{GSR}$) in the panels of Figure~\ref{fig:thetaGC}, to ascertain any plausible trends.  Each panel contains interesting information.  For example, the upper left panel shows that those post-SDSS satellites with high ellipticities ($\epsilon \gtrsim 0.4$) tend to have $20^{\circ} \lesssim \Delta \theta_{GC} \lesssim 40^{\circ}$.  This is roughly consistent with the broad peak in orientations witnessed by \citet{Klimentowski09}.  From the upper right panel of Figure~\ref{fig:thetaGC}, we find that the faintest satellites (which are also the nearest, due to selection effects) are the most aligned with the MW center, with $\Delta \theta_{GC}$ centered around $\sim20^{\circ}-30^{\circ}$ for those satellites with $M_{V} > -4.5$ mag (with the exception of Segue~2).   There is no clear correlation between  $\Delta \theta_{GC}$ and $M_{V}$ among the new MW satellites, however. 
The Spearman rank-order correlation coefficient for the new MW satellite sample does not indicate a true correlation between $ \Delta \theta_{GC}$ and $M_{V}$, although if Segue~2 is struck from the sample there is only a $\sim$4\% probability that the data are uncorrelated.

There are no discernible trends in $\Delta \theta_{GC}$ as a function of distance or $V_{GSR}$, which argues mildly against the idea that the observed satellite orientations have a tidal origin.  At apocenter, for instance, when the satellite is both distant and at rest with respect to the MW, one might expect the satellite to be most oriented toward the MW center.

To summarize, there is some correspondence between a subset of the new MW satellites' orientation -- particularly the faintest  ($M_{V} > -4.5$) and most elliptical ($\epsilon \gtrsim 0.4$) systems -- and the predictions from \citet{Klimentowski09}, suggesting that these systems are being affected by the tidal forces of the MW.  While these results are intriguing, they are difficult to interpret because of line of sight effects, the large uncertainties in current position angle measurements and the difficulty in comparing simulations and observations in this instance.  

\section{Summary and Conclusions}\label{sec:conclude}

We have utilized deep and wide field imaging of three of the most distant, faint and recently discovered MW satellites -- Leo~V, Pisces~II and CVn~II -- to study their SFH and structure.  Our data reached between $\sim$0.5 and 1 magnitudes below the main sequence turnoff, and our imaging fields of view were a factor of $\sim$10 larger than the satellites' half light radii.  In their structural properties and SFH, all three objects are relatively similar, with half light radii between $60 \lesssim r_{h} \lesssim 90$ pc and stellar populations which are old ($>10$ Gyr) and metal poor ([Fe/H]$\sim-2$).  However, Leo~V (and to a lesser extent CVn~II) shows direct signs of tidal disturbance due to nearby stream-like stellar overdensities and a spectroscopic [Fe/H] that is high for its luminosity.  Leo~V's Galactic alignment  is comparable to simulations of tidally disturbed dwarf galaxies.   Further deep, high-resolution imaging and wide-field kinematics of Leo~V is necessary to definitively determine its nature.

While all of the other MW satellites with $D>150$ kpc display clear signs of an extended SFH, the three objects in this paper do not.  This difference may indicate  that current MW distance alone is not a good indicator of SFH, and that other  factors -- such as orbital/infall history and initial baryonic reservoir -- may also play a role.  However, the low luminosities of Leo~V, Pisces~II, and CVn~II prevent us from completely ruling out a small amount of star formation in the past 10 Gyr, so we can not yet draw robust conclusions.

Most of the new MW satellites have now been followed up with deeper, wide-field imaging, making it an opportune time to study their properties as a population and in comparison to the classical dSphs.  We have collected the structural properties of the post-SDSS MW satellites into Table~\ref{table:structparam}, along with their classical dSph counterparts.  Unlike previous work, which had only uniform access to the shallow SDSS discovery data itself, there is no evidence that the SDSS satellites and the classical dSphs have a different ellipticity distribution as a population (at the $\sim$2-$\sigma$ level), even though individual objects have extreme ellipticities.  

Finally, we have investigated the major axis orientation of the satellites with respect to the MW center, $\Delta \theta_{GC}$.  Intriguingly, the nearest and faintest ultra-faint satellites tend to be most aligned with the MW center.  Additionally, post-SDSS satellites with $\epsilon \gtrsim 0.4$ tend to have $20^{\circ} \lesssim \Delta \theta_{GC} \lesssim 40^{\circ}$, comparable to numerical simulations of dwarf galaxies undergoing tidal stirring in a MW-like halo \citep{Klimentowski09}.
Although tentative, we may be seeing the imprint of tidal forces on the faintest satellites.

\acknowledgements

Many thanks to Maureen Conroy and John Roll for their tireless efforts related to Megacam, and their focus on getting scientific results.  Based in part on data collected at Subaru Telescope and obtained from the SMOKA, which is operated by the Astronomy Data Center, National Astronomical Observatory of Japan.  BW acknowledges financial support from NSF grant AST-0908193 for support.  EO acknowledges partial support from NSF grant AST-0807498.  DZ acknowledges financial support from NSF grant AST-0907771.

\bibliographystyle{apj}
\bibliography{apj-jour,mybib}

\clearpage

\begin{deluxetable*}{lcccccccccc}
\tablecolumns{10}
\tablecaption{Summary of Observations and Completeness by Field \label{table:observations}}
\tablehead{
\colhead{Dwarf Name}  & \colhead{Telescope/} & \colhead{UT Date} &
 \colhead{$\alpha$} & \colhead{$\delta$} & \colhead{Filter} &
 \colhead{Exposure} & \colhead{PSF FWHM} &\colhead{50\%} & \colhead{90\%}\\
\colhead{} & \colhead{Instrument}
 &\colhead{}&\colhead{(J2000.0)}&\colhead{(J2000.0)}&\colhead{}&\colhead{Time
 (s)}&\colhead{(arcsec)} &\colhead{Comp (mag)} & \colhead{Comp (mag)}}\\
\startdata

Pisces~II & Clay/Megacam & 2010 Oct 03 & 22:58:31.0 & +05:57:09.0 & $g$ &
 $9\times 270$ & 1.1 & 25.75 & 24.55\\
& & & & & $r$ & $6\times 300$ & 0.9 & 25.30 & 23.80 \\
Leo~V & Clay/Megacam & 2010 April 13 & 11:31:09.0 & +02:13:12.0 & $g$ &
 $6\times 300$ & 0.7 & 26.20 & 25.30 \\
& & & & & $r$ & $6\times 300$ & 0.6 & 25.75 & 24.70\\
CVn~II & Subaru/Suprimecam & 2008 April 3,5 & 12:57:10.0 & +34:19:15.0 & $V$ &$10\times120$ & 1.1 & 25.50 &24.30\\
& & & & & $I$ & $15\times200$ & 0.9 & 25.00& 23.60\\
\enddata

\end{deluxetable*}

\clearpage

\begin{deluxetable*}{lcccccccccc}
\tablecolumns{11}
\tablecaption{Leo V Photometry \label{table:phot_Leo V}}
\tablehead{
\colhead{Star No.} & \colhead{$\alpha$} & \colhead{$\delta$} & \colhead{$g$} &\colhead{$\delta g$} & \colhead{$A_{g}$} & \colhead{$r$} &\colhead{$\delta r$} & \colhead{$A_{r}$} &\colhead{SDSS or Mag}\\
\colhead{} & \colhead{(deg J2000.0)} & \colhead{(deg J2000.0)} & \colhead{(mag)}&\colhead{(mag)} & \colhead{(mag)} & \colhead{(mag)} & \colhead{(mag)} & \colhead{(mag)} &\colhead{}}
\startdata
0&172.83395&2.197776&15.89&0.01&0.10&15.51&0.01&0.07&SDSS\\
1&172.82309&2.256493&17.18&0.02&0.10&16.58&0.01&0.07&SDSS\\
2&172.85682&2.194839&17.34&0.01&0.10&16.89&0.01&0.07&SDSS\\
3&172.85827&2.206850&17.35&0.01&0.10&16.68&0.01&0.07&SDSS\\
4&172.84439&2.194516&17.96&0.01&0.10&17.13&0.01&0.07&SDSS\\
5&172.86004&2.186176&19.33&0.01&0.10&18.37&0.01&0.07&SDSS\\
6&172.73155&2.250007&17.37&0.02&0.10&16.96&0.01&0.07&SDSS\\
7&172.73697&2.255144&18.27&0.02&0.10&17.49&0.01&0.07&SDSS\\
8&172.84020&2.263564&18.05&0.02&0.10&16.61&0.01&0.07&SDSS\\
9&172.77737&2.136055&16.09&0.02&0.10&15.39&0.01&0.07&SDSS\\
\enddata
\tablenotetext{a}{See electronic edition for complete data table.}
\end{deluxetable*}

\clearpage

\begin{deluxetable*}{lcccccccccc}
\tablecolumns{11}
\tablecaption{Pisces II Photometry \label{table:phot_Pisces II}}
\tablehead{
\colhead{Star No.} & \colhead{$\alpha$} & \colhead{$\delta$} & \colhead{$g$} &\colhead{$\delta g$} & \colhead{$A_{g}$} & \colhead{$r$} &\colhead{$\delta r$} & \colhead{$A_{r}$} &\colhead{SDSS or Mag}\\
\colhead{} & \colhead{(deg J2000.0)} & \colhead{(deg J2000.0)} & \colhead{(mag)}&\colhead{(mag)} & \colhead{(mag)} & \colhead{(mag)} & \colhead{(mag)} & \colhead{(mag)} &\colhead{}}
\startdata
0&344.59997&5.958982&18.56&0.02&0.25&18.07&0.02&0.18&SDSS\\
1&344.60326&5.965916&16.99&0.02&0.25&15.68&0.02&0.18&SDSS\\
2&344.61875&5.971154&17.00&0.02&0.24&16.57&0.02&0.18&SDSS\\
3&344.65301&5.954814&16.88&0.02&0.24&16.05&0.02&0.17&SDSS\\
4&344.62868&5.929240&19.06&0.02&0.24&18.01&0.02&0.17&SDSS\\
5&344.64381&5.998345&17.43&0.02&0.24&16.05&0.01&0.17&SDSS\\
6&344.66289&5.956607&16.34&0.02&0.24&15.91&0.02&0.17&SDSS\\
7&344.66465&5.958012&17.34&0.02&0.24&16.64&0.02&0.17&SDSS\\
8&344.59092&5.947631&19.53&0.02&0.25&18.10&0.02&0.18&SDSS\\
9&344.61120&5.926302&19.27&0.02&0.24&18.15&0.02&0.18&SDSS\\
\enddata
\tablenotetext{a}{See electronic edition for complete data table.}
\end{deluxetable*}

\clearpage

\begin{deluxetable*}{lcccccccccc}
\tablecolumns{11}
\tablecaption{CVn II Photometry \label{table:phot_CVn II}}
\tablehead{
\colhead{Star No.} & \colhead{$\alpha$} & \colhead{$\delta$} & \colhead{$V$} &\colhead{$\delta V$} & \colhead{$A_{V}$} & \colhead{$I$} &\colhead{$\delta I$} & \colhead{$A_{I}$} &\colhead{SDSS or Sub}\\
\colhead{} & \colhead{(deg J2000.0)} & \colhead{(deg J2000.0)} & \colhead{(mag)}&\colhead{(mag)} & \colhead{(mag)} & \colhead{(mag)} & \colhead{(mag)} & \colhead{(mag)} &\colhead{}}
\startdata
0&194.28736&34.31227&19.09&0.02&0.03&16.00&0.02&0.01&SDSS\\
1&194.26727&34.32405&19.92&0.03&0.03&18.87&0.03&0.01&SDSS\\
2&194.29775&34.31017&20.48&0.04&0.03&18.94&0.03&0.01&SDSS\\
3&194.27426&34.29918&18.62&0.02&0.03&17.77&0.02&0.01&SDSS\\
4&194.30703&34.32029&19.12&0.02&0.03&16.82&0.02&0.01&SDSS\\
5&194.30683&34.31303&19.28&0.02&0.03&18.10&0.02&0.01&SDSS\\
6&194.30428&34.33148&20.65&0.04&0.03&18.57&0.02&0.01&SDSS\\
7&194.26234&34.30734&21.94&0.14&0.03&18.36&0.02&0.01&SDSS\\
8&194.28318&34.28882&16.77&0.02&0.03&16.06&0.02&0.01&SDSS\\
9&194.23909&34.32260&17.26&0.02&0.03&15.49&0.02&0.01&SDSS\\
\enddata
\tablenotetext{a}{See electronic edition for complete data table.}
\end{deluxetable*}

\clearpage

\begin{deluxetable*}{lcccccc}
\tablecolumns{7}
\tablecaption{Horizontal Branch Distance Measurements at Different Metallicities  \label{table:distances}}
\tablehead{
\colhead{}  & \multicolumn{2}{c}{M92 ([Fe/H]=$-$2.4)} &  \multicolumn{2}{c}{M3/M13 ([Fe/H]=$-$1.5)} &  \multicolumn{2}{c}{Adopted ([Fe/H]=$-$2.0)}\\
\colhead{Satellite} & \colhead{Distance (kpc)} & \colhead{$(m-M)_0$} & \colhead{Distance (kpc)} & \colhead{$(m-M)_0$} & \colhead{Distance (kpc)} & \colhead{$(m-M)_0$} } \\
\startdata
Pisces~II & $187\pm10$ & $21.36\pm0.12$ &$175\pm12$ & $21.22\pm0.15$ & $183\pm15$ & $21.31\pm0.17$\\
Leo~V &  $194\pm11$& $21.44\pm0.13$ & $187\pm13$ & $21.36\pm0.15$ &$196\pm15$ & $21.46\pm0.16$ \\
\enddata
\tablenotetext{}{Columns 1 and 2 are HB-derived distances assuming M92 and M3/M13 as fiducials, respectively.  Our adopted distance, in column 3, is the M3/M13 distance corrected to an assumed metallicity of [Fe/H]=$-$2. See \S~\ref{sec:dist} for details.  }
\end{deluxetable*}

\clearpage

\begin{deluxetable*}{lcccccc}
\tabletypesize{\small}
\tablecolumns{10}
\tablecaption{Parameterized fits to Leo~V, Pisces~II and CVn~II \label{table:paramfits}}

\tablehead{
\colhead{Parameter} & \colhead{Measured} & \colhead{Uncertainty} & \colhead{Measured} & \colhead{Uncertainty} & \colhead{Measured} & \colhead{Uncertainty}\\
\colhead{} & \colhead{Leo~V} & \colhead{} & \colhead{Pisces~II} & \colhead{} & \colhead{CVn~II}\\
}
\startdata

$M_{V}$ & $-4.4$ & $\pm0.4$& $-4.1$&$\pm0.4$ & $-4.6$ & $\pm0.2$ \\
$D$ (kpc) &  196& $\pm$15 & 183 & $\pm$15 & 160\tablenotemark{a} & $\pm$7\\ 
\hline
Exponential Profile \\
\hline\hline
RA (h~m~s)&11:31:08.17&$\pm4''$&22:58:32.33 & $\pm5"$ & 12:57:10.04 & $\pm4"$\\
DEC (d~m~s)&+02:13:19.38&$\pm2''$&+05:57:17.7 & $\pm4"$ & +34:19:14.39 & $\pm5"$\\
$r_{h}$ (arcmin) &1.14&$\pm0.53$&1.09 & $\pm$0.19 & 1.83 & $\pm$0.21\\
(pc) & 65 & $\pm30$ & 58 & $\pm10$ & 85 & $\pm$10 \\
$\epsilon$&0.52&$\pm0.26$&$<0.28$\tablenotemark{b} & ... & 0.39 & $\pm$0.07 \\
$\theta$ (degrees)&90.0&$\pm$10.0&107 & Unconstrained & -5.8 & $\pm$8.0\\
\hline
Plummer Profile \\
\hline\hline
RA (h~m~s)&11:31:08.08&$\pm5''$&22:58:32.20 & $\pm5"$ & 12:57:09.98 & $\pm4"$\\
DEC (d~m~s)&+02:13:19.47&$\pm2''$&+05:57:16.3 & $\pm4"$ & +34:19:16.97 & $\pm6"$\\
$r_{h}$ (arcmin) &0.95&$\pm0.47$&1.12 & $\pm$0.18 & 1.86 & $\pm$0.18 \\
(pc) &54&$\pm27$&60 & $\pm10$ & 86.6 & $\pm$8.4\\
$\epsilon$&0.52& $\pm0.18$&0.33 & $\pm0.13$ & 0.37 & $\pm0.06$\\
$\theta$ (degrees)&90.7&$\pm10.1$&110&$\pm11$ & -6.9 & $\pm7.0$\\
\hline\hline

\enddata \tablenotetext{a}{The distance to CVn~II was taken from the RR Lyrae study of \citet{Greco08}}
\tablenotetext{b}{Here $\epsilon$ corresponds to the 68\% upper confidence limit, given that our derived $\epsilon$ is consistent with 0.0.}
\end{deluxetable*}

\clearpage

\begin{deluxetable*}{lccccccccc}
\tablecolumns{9}
\tablecaption{ MW satellite parameters \label{table:structparam}}
\tablehead{
\colhead{Satellite}  & \colhead{$M_{V}$} &\colhead{Dist} &  \colhead{$R_{half}$} & \colhead{Ellipticity} & \colhead{Pos. Angle} & \colhead{$V_{GSR}$} & \colhead{$\Delta \theta_{GC}$} & \colhead{Ref.}\\
\colhead{} &\colhead{(mag)} & \colhead{(kpc)} & \colhead{(pc)} & \colhead{} & \colhead{(degrees)}  & \colhead{km s$^{-1}$} & \colhead{(degrees)} & \colhead{}}\\
\startdata
Hercules &  $-6.2\pm0.4$ & $133\pm6$&  $229\pm19$ & $0.67\pm0.03$ & $-72.36\pm1.65$ & $142.9\pm1.1$& 49.3& 1,16 \\
Bootes I & $-6.3\pm0.2$ & $66\pm3$ & $242\pm21$& $0.39\pm0.06$ & $14\pm6$&$109.4\pm0.7$ & 65.0&8,12,17 \\
Bootes II & $-2.2\pm0.7$ & $42\pm2$ & $31\pm10$&$0.27\pm0.15$ & $-33\pm57$&$-115.6$\tablenotemark{b} &18.7 & 2,18 \\ 
Leo IV & $-5.5\pm0.3$ & $154\pm5$& $128\pm29$ & $<0.23$ & Unconstrained &$14.0\pm1.4$ & NA & 3,4,16 \\
Leo~V & $-4.4\pm0.4$&  $196\pm15$ &  $70.9\pm27.6$ &$0.55\pm0.22$ & $89.2\pm9.0$ & $63.0$\tablenotemark{c} & 29.7 &14,19\\
Pisces~II &  $-4.1\pm0.4$& $183\pm15$ &$57.2\pm10.0$ & $<0.31$ & Unconstrained& Unknown & NA &14\\
CVn~II &  $-4.6\pm0.2$ & $160\pm7$ &  $85.2\pm9.8$ & $0.39\pm0.07$ & $-5.8\pm 8.0$ & $-96.1\pm1.2$ & 50.7 &14,16\\
CVn~I &$-8.6^{+0.2}_{-0.1}$ &  $218\pm10$ & $564\pm36$ & $0.39\pm0.03$ & $70^{+3}_{-4}$ & $76.8\pm0.6$ & 57.9 &8,15,16\\
Ursa Major I & $-5.5\pm0.3$ &  $96.8\pm4$ & $318^{+50}_{-39}$ & $0.80\pm0.04$& $71^{+2}_{-3}$& $-8.9\pm1.4$& 25.1 &8,34 \\
Ursa Major II & $-4.0\pm0.6$  & $30\pm5$& $123\pm3$ & $0.50\pm0.02$ & $-74.8\pm1.7$ & $-36.5\pm1.9$ & 41.9 & 5,6,17 \\
Coma Berenices & $-3.8\pm0.6$  &$44\pm4$&$74\pm4$ & $0.36\pm0.04$ & $-67.0\pm3.6$& $82.6\pm0.9$ & 7.5 & 5,7,17 \\
Willman~1 &$-2.7\pm0.7$  & $38\pm7$ &  $25\pm6$ & $0.47^{+0.07}_{-0.08}$&$77\pm5$& $-49.3\pm1.0$ & 22.3 & 8,13,20\\
Segue~1 &$-1.5^{+0.6}_{-0.8}$& $23\pm2$ &$29^{+8}_{-5}$&$0.48^{0.10}_{-0.13}$&$85\pm8$&$116.1\pm0.9$ & 29.7 &21\\
Segue~2  & $-2.5\pm0.2$&35&$34\pm2$&$0.15\pm0.1$&$182\pm17$&Unknown& 66.7 & 33\\
\hline
Classical Sats \\
\hline\hline
Draco & $-8.75\pm0.15$ & $76\pm5$ &  $221\pm16$ & $0.31\pm0.02$ & $89\pm2$ & -101.6\tablenotemark{d} & 85.4 & 8,9,22\\
Fornax & $-13.3^{+0.2}_{-0.3}$ &  $147\pm3$ &$714\pm40$& $0.30\pm0.01$ & $41\pm1$& $-29.2\pm0.1$  & 0.5 & 11,23,25,26,27\\
Sculptor &$-11.2^{+0.3}_{-0.5}$&$86\pm5$ &$282\pm41$ & $0.32\pm0.03$ & $99\pm1$& $83.3\pm0.1$ & 43.0 & 11,23,25,26,28 \\
Carina &   $-9.3\pm0.2$& $105\pm2$ &$254\pm28$&  $0.33\pm0.05$& $65\pm5$ & $13.5\pm0.1$& 77.4 &  11,23,25,26,27 \\
Leo I & $-11.9\pm^{+0.3}_{-0.4}$ &  $254\pm18$& $295\pm49$ & $0.21\pm0.03$ & $79\pm3$& $177.8\pm0.5$ & 37.5 &10,11,24,25,26\\
Sextans &$-9.6^{+0.3}_{-0.4}$ & $96\pm3$&$768\pm47$ & $0.35\pm0.05$ & $56\pm5$ & $80.3\pm0.1$& 65.6 &11,23,25,26\\
Leo~II  & $-10.0^{+0.4}_{-0.3}$&$233\pm15$  &$177\pm13$ & $0.13\pm0.05$ & $12\pm10$& $26.6\pm0.6$ & 77.0 & 11,26,30,31 \\
Ursa~Minor & $-9.2\pm0.4$& $77\pm4$  & $445\pm44$&$0.56\pm0.05$ & $53\pm5$ & $-87.6\pm2.0$& 87.2 & 11,22,25,26,32\\
\enddata \tablenotetext{a}{References: (1) \citet{sandherc}, (2) \citet{Walsh08}, (3) \citet{Moretti09}(4) \citet{Sandleoiv}, (5) \citet{Munoz10}, (6) \citet{Zucker06}, (7) \citet{Belokurov07}, (8) \citet{sdssstruct}, (9) \citet{Bonanos04}, (10) \citet{Bellazzini04}, (11) \citet{Irwin95}, (12) \citet{dallora06},(13) \citet{Willman06} (14) This work, (15) \citet{Martin08}, (16) \citet{simongeha}, (17) \citet{Koposov11}, (18) \citet{Koch09}, (19) \citet{leovspec}, (20) \citet{Willman11}, (21) \citet{Simon11} ,(22) \citet{Munoz05}, (23) \citet{Walker09EM}, (24) \citet{M08}, (25) \citet{Mateo98}, (26) \citet{Wolf10} , (27) \citet{Pietrzynski09}, (28) \citet{Pietrzynski08}, (29) \citet{Lee03}, (30) \citet{Coleman07}, (31) \citet{Bellazzini05}, (32) \citet{Carrera02}, (33) \citet{Belokurov09}, (34) \citet{Okamoto08} }
\tablenotetext{b}{The Bootes II systemic velocity measurement of \citet{Koch09} did not include an uncertainty}
\tablenotetext{c}{The Leo~V systemic velocity is an estimate based on Figure~2 of \citet{Walker09}.}
\end{deluxetable*}

\clearpage

\clearpage

\clearpage

\clearpage

\clearpage

\clearpage

\clearpage

\end{document}